\documentclass[prd,preprint,eqsecnum,nofootinbib,amsmath,amssymb,
               tightenlines,dvips]{revtex4}

\usepackage{graphicx}
\usepackage{bm}
\newcommand{\be}{\begin{eqnarray}}
\newcommand{\ee}{\end{eqnarray}}
\newcommand{\bea}{\left (\begin{array}{cc}}
\newcommand{\eea}{\right )\end{array}}

\newcommand\del{\partial}

\newcommand{\mat}{\left ( \begin{array}{cc}}
\newcommand{\emat}{\end{array} \right )}
\newcommand{\matt}{\left ( \begin{array}{ccc}}
\newcommand{\ematt}{\end{array} \right )}
\newcommand{\matf}{\left ( \begin{array}{cccc}}
\newcommand{\ematf}{\end{array} \right )}
\newcommand{\vect}{\left ( \begin{array}{c}}
\newcommand{\evect}{\end{array} \right )}

\def\eq#1{Eq.~(\ref{#1})}

\begin{document}
\title{Hydrodynamic Flow from Fast Particles}

\author{J. Casalderrey-Solana, E. V. Shuryak, D. Teaney}
\affiliation
   {
    Department of Physics \& Astronomy,
    SUNY at Stony Brook,
    Stony Brook, NY 11764, USA
   }

\date{\today}

\begin{abstract}
We study the interaction of a fast moving particle in the 
Quark Gluon Plasma with linearized hydrodynamics.  
We derive the linearized 
hydrodynamic equations on top of an expanding fireball, 
and detail the
solutions for 
a static medium. There are two modes far from the jet -- a sound mode and 
a diffusion mode. 
The diffusion mode 
is localized in a narrow wake behind the jet while the sound 
mode propagates at the Mach angle, $\cos(\theta_M) = c_s/c$. 
A general argument
shows that the strength
of the diffusion mode relative to the sound
mode is directly  proportional to the entropy produced 
by the jet-medium interaction. This argument  does 
not rely on the linearized approximation and the  
assumption of local thermal equilibrium close to the jet.  
With this insight we calculate the
spectrum of secondaries associated with the fast moving particle.
If the energy loss is large and the jet-medium interaction
does not produce significant entropy, the flow at the 
Mach angle can be observed in the associated spectrum.
However, the shape of associated spectra
is quite fragile and sensitive to many of the inputs of the calculation.
\end{abstract}

\maketitle

\section{Introduction}
One of the major findings at RHIC is jet quenching,  the suppression
of high transverse momentum particles \cite{jetquenching}.  This strong
suppression is attributed to the energy  loss
of partons traversing the dense medium formed 
in a high energy heavy ion collision. 
Different microscopic mechanisms for the jet energy loss
have been proposed \cite{early,Gyulassy_losses,Kovner:2003zj,Dok_etal,SZ_dedx} 
(see Ref.~\cite{xnwang_workshop} for a review).

Ultimately  the energy ``lost" by
these fast  partons 
is shared among the constituents of the medium.
In a previous paper  we  argued that 
the fate of the deposited energy and momentum should be described by
hydrodynamics\footnote{see Ref.~\cite{Stocker} for a similar idea by Stocker.} \cite{CST}. Hydrodynamic models describe the RHIC data 
reasonably well and motivated this suggestion.  
The suggestion was also motivated by the  
experimental observation 
of particles associated with a high $p_T$ trigger at
a particular azimuthal angle, $\Delta \phi = \pi - 1.2~\mbox{rad}$,
relative to the trigger particle \cite{phenix_peaks, star_peaks}.
Within linearized hydrodynamics, jets induce
flow fields with a conical shape similar to the supersonic flow past planes.
It was argued that this conical flow could provide
an explanation for the observed correlation.
Discussion of alternative explanations to the large
$p_T$ correlations at 
RHIC can be found in Refs.~\cite{RM,Dremin,MWK,Vitev}.

In this first paper we solved linearized 
hydrodynamics in a 
static homogeneous medium, and
found that 
two modes can be excited -- a sound mode  
and diffusion mode \cite{CST}. The relative 
strength of these modes depends on the jet-medium interaction 
and will be  clarified in this work.
Subsequently, the effect 
of the fireball expansion 
was  studied \cite{S_w}, and  
an attempt to incorporate more realistic mediums has been made 
\cite{Renk}.
Further, the  nonlinear hydrodynamic response to the jet was estimated  with 2+1 dimensional hydro code \cite{Heinz}.
The conical flow was not observed in the azimuthal correlations
of that work; possible reasons  for this are discussed in section \ref{nad}. 
Finally, a parton transport calculation based on the AMPT 
model also found large angle di-hadron correlations \cite{AMPT, MAMPT}.

In a second paper two of us 
studied  the conical flow in an expanding
fireball with a variable speed of sound and found that 
the amplitude of the sound wave increases \cite{CS}.
 Furthermore, we have shown that a first order
phase transition with a vanishing speed of sound, would lead to more
peaks in the di-hadron correlation function. These 
additional peaks have not been observed.

In this paper we present a systematic study of the conical flow within the linearized theory. In section \ref{setting}
we start by providing an overview and 
explaining the approximations.
Then in section \ref{LH}, we review our
calculation  for a static medium and find the 
two modes of linearized hydrodynamics \cite{CST}.
In section \ref{pot}, we describe  
potential flow, first for an ideal fluid, and then with viscosity.
Section \ref{pot} also provides
the linearized equations which propagate small perturbations 
over a general hydrodynamical expansion. 
In section \ref{EL},
we relate the energy and momentum loss of the 
jet to the outgoing sound waves at
large distances.  This result 
ties the initial energy loss  to the normalization of the final
spectrum of secondaries that is calculated in section \ref{NSP}.
Our work in section \ref{EL} also delineates
two microscopic models for the jet-medium interaction 
based on entropy production.
These two models are described in
sections \ref{ad} and \ref{nad}, and their corresponding 
correlation functions are calculated in section \ref{NSP}. Finally, 
many technical derivations are delegated into appendices.

\section{\label{setting} Overview}

\begin{figure}
\begin{center}
 \includegraphics[width=7cm]{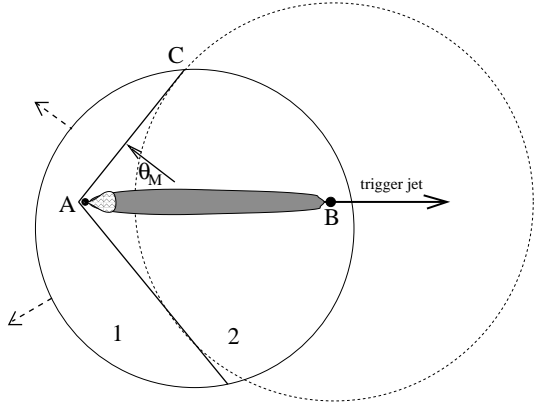}
 \caption{\label{fig_shocks}
A schematic picture of the flow created by a jet moving through
the fireball. The trigger jet is moving to the right
away from the origination point (the black circle at point B).
Sound waves start propagating as spherical waves (the
dashed circle) from the origination point. The
  companion quenched jet is moving to the left
creating a wake of matter (shaded area) and adding to the sound wave.
 The head of the jet is a non-equilibrium gluonic
shower formed by the original hard parton (black dot A).
The solid arrow indicates the flow velocity which is perpendicular to the shock cone at
the angle $\theta_M$, $\cos(\theta_M) = c_s/c \simeq 0.55$. 
}
\end{center}
 \end{figure}

A schematic picture of di-jet production in a 
nucleus-nucleus collision is given in Fig.~\ref{fig_shocks}.
For simplicity we will only consider a di-jet pair with
both particles at mid-rapidity in a central Au-Au collision. One
jet serves as the trigger particle and is biased toward the 
edge of the nucleus due to the strong jet quenching. The 
other jet  then travels a distance $2R_{\rm Au} \sim 10\,\mbox{fm}$
and is either completely or partly absorbed. Experimentally 
a companion jet with $p_{T} \sim 10\,\mbox{GeV}$ seems to 
be absorbed while a jet with larger transverse momentum seems
to ``punch through'' and re-appear at an angle $\Delta \phi = \pi$ relative to the trigger jet.

In general the rapidity of the two jets are not identical and this will
complicate the experimental interpretation of our results.  The
particular configuration we are considering can be selected with three
particle correlations and currently there is an experimental effort in
this direction.

Far from the jet we can calculate the correlation of $T^{\mu\nu}$ 
with the passage of the jet using linearized hydrodynamics. 
Generally there are two modes -- a sound mode and a diffusion
mode. The diffusion mode is concentrated
in a narrow wake behind the jet while the sound mode 
propagates forward at the Mach angle, $\cos(\theta_{M}) = c_s/c \simeq 0.55$.
The total momentum loss can be related to the amplitude of 
the wake and the amplitude of the sound wave. 
Under the reasonable assumption that the total energy and momentum
loss of the jet are equal we reach the stronger conclusion 
which relates the entropy production of the jet to the 
relative strength of the two modes. 

The equations for the sound wave are given by Eqs.~(\ref{def_pot}), and (\ref{phi_fluid}). The unknown function in this formula, $dF/dx$,  is
related to the momentum transfered to the sound wave -- see \eq{dedt}. 
Similarly the flow fields in the wake are given by 
Eq.~(\ref{Rgaus}), and the amplitude $A$ in this formula is 
fixed by the momentum transfered to the wake -- see \eq{Pwake}.
This momentum transfer is in turn related by \eq{Swake} 
to the total entropy
produced by the jet. In summary, by specifying 
the total rate of energy loss and entropy production the flow 
fields at large distances are determined.

We next estimate the 
 medium modifications at the  head of the jet where
the jet loses energy through collisional and radiative processes. 
 Since the jet energy
is large  compared to the typical scale of the medium, the jet acts as 
a point source. The energy deposited by this source is  distributed
into the surrounding fluid through 
highly dissipative 
processes. 
To estimate the order of magnitude of the initial 
modification,  consider a jet that losses certain amount of energy per 
unit length,
$dE/dx$. This energy is absorbed  by the medium
over a dissipative length scale
of the order of the sound attenuation length,  
$\Gamma_s=4\eta/3w$, where $\eta$ is the shear viscosity 
and $w=e+p$ is enthalpy of the medium. Thus 
we
compare the energy deposited by the jet in this typical length
$\Gamma_s$ with the energy of the
fluid in a deposition volume of $\Gamma_s^3$,
\be
\frac{E_{\rm lost}}{E_{\rm fluid}}\approx 
                       \frac{\frac{dE}{dx}\times\Gamma_s}{e\times\Gamma_s^3}		      \, .
\ee

To estimate this initial modification, we  
use the formula from Ref.~\cite{Dok_etal} for radiative energy loss 
\be
-\frac{dE}{dx}=\frac{\alpha_s C_R}{8}\,\frac{\mu^2}{\lambda_g}\, L 
\ln \frac{L}{\lambda_g} \, ,
\ee
where $\frac{\mu^2}{\lambda_g}=\hat q$ is the transport coefficient. 
Perturbative estimates for this parameter 
give $\hat q=0.6 ~\mbox{GeV}^2/\mbox{fm}$ 
for $T=200~\mbox{MeV}$ \cite{Baier}. Setting 
$\lambda_g \approx 1~ \mbox{fm}$ \cite{Baier},
$L=5 \mbox{fm}$, and $C_R=3$ (gluons), we obtain, $dE/dx \approx 2.7 ~\mbox{GeV/fm}$.
Similarly, perturbative estimates  for the sound 
attenuation length give with $\alpha_s=1/2$, $\Gamma_s\approx 0.18/T$ \cite{Baymetal,AMY6}.
The conjectured lower bound for $\Gamma_s$ is $1/3\pi T$ \cite{visc_bound}. 
 For the energy density of
the unperturbed medium ($e$) we will take the QGP value 
as measured on the lattice,
 $e\approx 12 T^4$ \cite{karsch}. For $T=200\,\mbox{MeV}$, we 
finally conclude that
\be
\frac{E_{\rm lost}}{E_{\rm fluid}}\approx 
                       \frac{\frac{dE}{dx}\times\Gamma_s}{e\times\Gamma_s^3}
		       \approx 36-100 \gg 1 \, .
\ee
(The range of values is  set by the range in $\Gamma_s$.) Thus, 
this quantity is numerically large although  
it is suppressed suppressed by $\alpha_s^7$ in perturbation
theory.  

 Since this number is greater than one, the 
jet is surrounded by its own small ``fireball''
(of size $\Gamma_s$ or more) where  variation 
of the thermodynamic quantities
 is very large and hydrodynamics cannot be applicable. 
Outside of this region, there is a domain where gradients are small enough
that viscous hydrodynamics can in principle be used, but
the behavior of the fluid is 
non-linear, dissipative, and possibly turbulent.
We will not discuss these complex regions in the present work. 

  Our current objective  is to study what happens 
 far from the jet, where
 the situation becomes less violent and more tractable.
Specifically,
it is the region where the flow velocity and pressure
 modifications
may be considered small compared the unperturbed medium, $i.e.$ 
 the region where {\em linearized} hydrodynamics can be used.
We note that 
it is possible to calculate the drag force on an airplane
using linearized hydrodynamics, by computing  the 
momentum flow through a large cylinder around the plane. 
The analogous computation for jets 
is performed in section~\ref{EL}.

In summary one may
 separate space-time into three regions
\begin{enumerate}
\item A region very close to the jet, $i.e.$  a region of size
 $\Gamma_s$,  where strong dissipative effects  happen.
\item A region  close to the jet where hydrodynamics is 
still nonlinear and possibly turbulent.
\item A region far from the jet where the perturbations and  gradients
are small. In this region one may use  linearized 
hydrodynamics and the motion is laminar.
\end{enumerate}

Since the primordial interaction of the jet with the 
fluid is complicated and non-linear, 
we will only study flows in the third linearized region,  
and  impose  only general constraints on the 
total energy, momentum, and entropy deposited by the jet.

\section{\label{LH} Linearized Hydrodynamics in a Static Medium}
In this section we will review our results of Ref.~\cite{CST} where we
 studied  small perturbations
of a homogeneous  relativistic baryon free fluid at 
rest.  The linearized hydrodynamic equations are written in
 terms
of the small quantities  
\be
\epsilon=\delta  T^{00} ~~,~~
g^i=\delta   T^{0i} ~.
\ee
The remaining components can be expressed in terms of these quantities by means 
of the equation of state. For a fluid with shear viscosity $\eta$
and vanishing bulk viscosity these remaining
components are
\be
T^{ij}=c^2_s \epsilon \delta^{ij}-\frac{\eta}{w} \left< \partial^i g^j \right >
\ee
where $c^2_s=dp/d\epsilon$ is the speed of sound of the fluid, $w$ the enthalpy
 and 
 $\left\langle \partial^i g^j \right\rangle $ is  symmetric traceless tensor 
\be
 \left \langle \partial^i g^j \right \rangle=\partial^i g^j+\partial^j g^i-\frac{2}{3}\delta^{ij}\partial_i g^i ~. 
\ee
By writing the energy and momentum conservation equations
 $\partial_\mu \delta T^{\mu \nu}=0$, the linearized hydro equations 
can be written after a spatial Fourier transforms as
\be
\label{eglsys}
\partial_t \epsilon+i k \,g_{\scriptstyle \rm L}=0 ~,\nonumber   \\ 
\partial_t g_l+ic^2_s k\,\epsilon+
\frac{4}{3}\frac{\eta}{\epsilon_0+p_0}\, k^2g_{\scriptstyle \rm L}=0 ~,
\\
\label{trans}
\partial_t {{\bf g}_{\scriptstyle \rm T}} + \frac{\eta}{\epsilon_0+p_0}k^2 { {\bf g}_{\scriptstyle \rm T}}=0 ~, 
\ee
where ${{\bf g}}=g_{\scriptstyle \rm L}\frac{{\bf k}}{k}+{ {\bf g}_{\scriptstyle \rm T}}$.
Thus the equations of linear hydro decouple into
 two different modes -- a sound mode (\eq{eglsys}) 
   and 
a  diffuson mode (\eq{trans}).
 The excitation of these
two modes depends on the initial conditions.

In order to study these initial condition, we considered the disturbance 
produced by an infinitesimal displacement of a particle moving along
the $\hat{x}$ direction with velocity v in a time interval $dt_0$ at time $t_0$.
 According
to the axial symmetry of the problem, the most general expression for 
the initial disturbance is
\be
\label{ini_cond}
\epsilon_{dt_0}(t=t_0,{\bf x})=e_0(x,r) \\ \nonumber
{\bf g}_{dt_0}(t=t_0,{\bf x})=g_0(x,r)\delta^{ix}
+{\bf \nabla} g_{1}(x,r) 
\ee 
The source functions $e_0(x,r)$ and $g_{1}(x,r)$ excite only
 the sound mode, while the remaining function
$g_0(x,r)$ excites the diffuson mode.
The particular value of these functions depends on the
interaction of the jet and the fluid in the near region. As argued in 
the introduction, it is difficult to find the exact functional
form for these sources and they are only constrained
by the total energy, momentum, and entropy deposited by the jet.

The hydrodynamic disturbance  due to these infinitesimal 
displacements are calculated as an integral of the sources
\be
T^{0\mu}(x)=\int d^3y \left(K^{\mu}_{e}\left(x-y\right) e_0(y)+
                        K^{\mu}_{g_0}\left(x-y\right) g_0(y)+
                        K^{\mu}_{g_1}\left(x-y\right) g_1(y)\right) ~,
\ee
where the three kernels $K^{\mu}_{e}$,
$K^{\mu}_{g_0}$ and $K^{\mu}_{g_1}$  are \cite{CST}.
\be
\label{kere}
K^{\mu}_{e}&=&\sum_{i=\pm}
                   \left( \frac{-1}{4\pi r} \partial_r P_i(\Gamma_s),
                   \partial^k\frac{c_s}{4\pi r}i P_i(\Gamma_s)\right) ~, 
\\
\label{kerg}
K^{\mu}_{g_0}&=&\sum_{i=\pm}
                    \left(\partial_x \frac{iP_i(\Gamma_s)}{4\pi c_sr} ,
                    \partial^k\partial_x\frac{-\int^r_0 P_i(\Gamma_s)}
                                                          {4\pi r}\right) \nonumber \\
&+ &
		    \left(0,\partial^k\partial_x\frac{2 \int^r_0 P(\frac{3}{2}\Gamma_s) }{4\pi r} 
                       +
             \delta^{kx}\frac{ e^{\frac{-r^2}{2\frac{3}{2}\Gamma_s t}}}{\left(2\pi\frac{3}{2}\Gamma_s t\right)^{3/2}}
             \right)
~,
\\
K^{\mu}_{g_1}&=&\sum_{i=\pm}
                   \left( \frac{i}{4\pi c_s r} \partial^2_r P_i(\Gamma_s),
                   \partial^k\frac{-1}{4\pi r}\partial_r P_i(\Gamma_s)\right) 
~.
\ee
Here we have defined the functions,
\be
P_{\pm}(\Gamma_s)=\frac{1}{\sqrt{2\pi\Gamma_st}}e^{-\frac{(r\pm c_st)^2}{2\Gamma_s t}} ~~,~~
P(\frac{3}{2}\Gamma_s)=\frac{1}{\sqrt{2\pi\frac{3}{2}\Gamma_st}}e^{-\frac{r^2}{2\frac{3}{2}\Gamma_s t}} ~. 
\ee

In what follows we will try to clarify further the meaning of the 
different modes as well as their excitations. We will start addressing
the problem in the case of inviscid fluid and we will show how to connect
to the viscous case. The results presented for dihadron azimuthal distribution
will be performed through these solutions.

\section{\label{pot}Relativistic potential flow}
In  nonrelativistic laminar flow
 it is customary to
introduce  a potential, so that the velocity can be expressed as a 
gradient,  ${\bf v}=\nabla \phi$. Such flows are clearly
irrotational as the ${\bf \nabla} \times {\bf v}$ vanishes. The conservation of circulation
for inviscid fluids guaranties that potential flow remains irrotational
throughout the evolution of the liquid. Sound waves, in particular
correspond to small potential disturbances over some given flow.

One can define
 a relativistic analogue of the potential flow for an ideal fluid
in which the vorticity of the fluid vanishes \cite{landau}. Let us first
recall the 
definition of the
the vorticity in the relativistic case
\cite{Taub}
\be 
\label{vorticity}
 \omega _{\sigma \tau}=\partial_{[\mu} u_{\nu]}h^{\mu}_{\sigma}h^{\nu}_{\tau},
\ee
where $h^{\mu \nu}= \eta^{\mu \nu}-u^{\mu}u^{\nu}$ is the 
projector into the space perpendicular to the velocity.
From this definition, it is simple to see that any flow of the form
\be
\label{def_pot}
f\,u_{\mu}=\partial_{\mu}\phi,
\ee
with $f$ some function, will be irrotational. The scalar field $\phi$, called the potential, is the 
relativistic analogue of the nonrelativistic potential flow. 

Thus, one can try to find solutions of the ideal relativistic hydrodynamics of
the form \eq{def_pot}. In fact, the function  $f$  can be determined by
 requiring that the projection of the energy momentum conservation equation
 into the 
space perpendicular to $u^{\mu}$ vanishes,
\be
\label{tranverse_eq}
h_{\mu \rho}\partial_{\nu}T^{\nu \rho}=0~.
\ee
In the case of an ideal baryon free fluid, these equations can be expressed as,
\be
s \left(u^{\nu} \del_\nu \left(T u_{\mu}\right) - \del_{\mu} T \right)=0,
\ee 
where $s$ and $T$ are the entropy density and temperature of the fluid.
It is clear then that, by setting $f=T$ a velocity field of the form 
of \eq{def_pot} is a solution of these equations \cite{landau}. 
Thus, imposing irrotationality of the flow leaves
 us only one equation to solve for the baryon free fluid,
 the entropy continuity equation.
\be
\label{entopy_continuity}
\partial_{\mu} (s u^{\mu})=0 ~.
\ee
This equation can be used to find a non-linear equation 
for the potential $\phi$
(see appendix \ref{din_eq}), which is equivalent to the ideal equations
of motion of an irrotational fluid.

Let us conclude by some examples of potential flow. Obviously, the static homogeneous 
baryon free fluid (which we will use in the rest of the paper) is potential. A slightly
less trivial example is the boost invariant Bjorken solution \cite{Bj}. 
In appendix \ref{Bj_pot} we show how to find this solution from the potential. Finally, 
as in the nonrelativistic case, sound waves are irrotational disturbances
of the hydrodynamic solution and are therefore potential. 


\subsection{\label{SPD} Small perturbations of a potential flow}
We now will study small perturbations over a given potential hydrodynamic
solution. 
Even though the interaction of the jet with
the background fluid  leads to very 
violent  flows in the near zone,
the small perturbation approximation is always valid far enough away
from the high energy particle.

In the region where the perturbation is small, we will have modifications
both from the thermodynamic quantities and from the velocity field, 
$T'$, $u'_{\mu}$.

Let us start with the case where the perturbation of the fluid does not
change the irrotational character. In this case, the modified hydrodynamic
fields can be described by a perturbation of the potential
\footnote{
 Here and in the rest of the paper, the modification
of the all the hydrodynamic fields but the potential 
will be denoted by $~'$. The modification of the potential 
denoted by $\varphi$.
}
\be
\phi \rightarrow \phi +\varphi~ .
\ee
Expanding the definition of the potential \eq{def_pot} to first order
we find
\be
\label{perturbed_T}
T'=u^{\mu}\partial_{\mu} \varphi~, \\ \nonumber
T u'_{\mu}=h_{\mu \nu}\del^{\nu} \varphi~,
\ee
with $h_{\mu \nu} = \eta^{\mu \nu} - u^{\mu}u^{\nu}$. 
Note that since $(u + u')^2=1$, we have $u \cdot u'=0$ to
first order in the perturbation.

In a general case, since the interaction with the high energy particle
is very complicated, the modified field
may not be irrotational.
However, one can still define a modified potential
$\varphi$ that describes part of the perturbation. In order to do so,
we can separate the perturbed velocity field $u'_{\mu}=U_{\mu}+R_{\mu}$,
 such
that $U_{\mu}$ and $T'$ are related to the potential $\varphi$ as in 
\eq{perturbed_T}. The rotational component $R$ is then the part of the
perturbation that cannot be expressed as \eq{perturbed_T} and, thus,
is linearly independent from $U$; as $U$ is orthogonal to $u$, 
so is $R$.

We now discuss the equations of motion for the perturbations. 
As we have defined $U$ such that it can be thought of as coming
from a potential, the perturbed transverse equation \eq{tranverse_eq}
depends only on R (and the background quantities)
\be
T~R^{\nu} \del_{\mu} \del_{\nu}\phi+\del^{\nu}\phi \del_{\nu}(T~R_{\mu})=0,
\ee
with $\phi$ the potential of the background medium.
 The remaining equation is the 
the entropy continuity equation for the perturbation
\be
\label{perturbed_entropy}
\partial_{\mu} (s u^{\mu})'=0~.
\ee
Using the definitions of the speed of sound for a baryon free fluid
$c^2_s=(s/T)\,dT/ds$ and the perturbations \eq{perturbed_T} we can re-express
\eq{perturbed_entropy} as a second order differential equation 
for $\varphi$:
\be
\del_{\mu} \left( \frac{s}{T} 
                  \left(\eta^{\mu \nu}+\left(\frac{1}{c^2_s}-1\right)u^{\mu}u^{\nu} \right) 
                   \del_{\nu} \varphi 
                   \right) + \partial_{\mu}\left(s R^{\mu}\right)=0~.
\ee
Let us mention that such second order differential equation can be expressed 
as the d'Alambertian of a scalar field
in a particular metric \cite{Bilic} that is dependent on the thermodynamic
and velocity fields of the background medium:
\be
\frac{1}{\sqrt{-G}}
\partial_{\mu} \left(\sqrt{-G} G^{\mu \nu} \partial_{\nu} \phi\right)=
 -\frac{1}{\sqrt{-G}}\del_{\mu}\left( s~R^{\mu}\right)\,, \\ \nonumber
G^{\mu \nu}=c_s \frac {T}{s} 
                \left( \eta^{\mu \nu}+(\frac{1}{c^2_s}-1)u^{\mu}u^{\nu}
           \right)\,,
\ee
where the rotational component acts as sound source.
\footnote{This emission (that vanishes for the
static case)  may be important in the final observation of the conical flow}
The reduction of the sound equation to a Klein-Gordon equation 
in a gravitational
background is well known in nonrelativistic fluids 
where $G^{\mu\nu}$ is known as the acoustic metric \cite{viser}.

Let us finally write these general equation for a simple case, the
static homogeneous medium.
In this case,
the equations of motion for the rotational and irrotational field decouple.
For the irrotational field, we find the standard wave equation
\be
\frac{1}{c^2_s}\partial^2_t \phi - \nabla^2 \phi=0\,.
\ee
For the rotational mode, we obtain a non propagating equation
\be
\del_0 R_{i}=0\,.
\ee
These two modes coincide with what we called sound and
 diffuson mode in 
\cite{CST} in the case where the viscosity of the medium $\eta$ vanishes.
 As also found there, only the sound mode propagates, while the
energy/momentum of the diffuson mode remains 
localized close to the position of their deposition. 
Thus, as far away from the source of the perturbation (the jet) the
modification of the hydrodynamical fields are small, we expect that only
potential flow should be found in this region, since 
perturbation can only arrive to long distances form the source due to
propagation.  Thus the  irrotational flow should be
concentrated in the region close to the jet path. 
This is also what happens in the case of flow past bodies
in nonrelativistic fluids.

\subsection{\label{VRPF} Viscosity and Potential Flow}
Our previous discussion of the
 potential flow was made  for inviscid fluids only, while
now we will study the effect of the shear viscosity.
We show below that in the case of static and
homogeneous background medium, one can still find a potential
solution that describes the system. For non-relativistic fluids this fact
has been shown in \cite{VPF}.

The energy momentum tensor for a liquid with non-vanishing
shear viscosity is
 \cite{landau}
\be
T^{\mu \nu}=w u^{\mu}u^{\nu} -\eta^{\mu \nu}p + 
           \eta \left <\nabla ^{\mu} u^{\nu} \right> ~,
\ee 
where $w$ is the enthalpy,  $p$ is the pressure,
 $\nabla_{\mu}=h_{\mu \nu} \partial^{\nu}$, and
\be \left\langle\nabla ^{\mu} u^{\nu} \right\rangle =\nabla ^{\mu} u^{\nu}+\nabla ^{\nu} u^{\mu}
-\frac{2}{3} h^{\mu \nu}\nabla_{\mu} u^{\mu} ~.\ee
In the case 
of a viscous fluid, the entropy equation is obtained by  projecting 
$u^{\mu}\partial_{\nu} T^{\mu \nu}$ and leads
to 
\be
\partial_{\mu} (s u^{\mu})= 
               \frac{\eta}{T} \left<\nabla^{\mu} u^{\nu}\right> \partial_{\nu} u_{\mu}~. 
\ee
For a static  or homogeneous  medium, there is no
 entropy production due to 
 viscosity. 
 The defining feature of the potential flow
is
 zero vorticity, {\it i.e.} the existence of 
solutions of the 
hydrodynamic equation in the form of \eq{def_pot}. 
For small perturbations on top a static fluid,
viscosity
will modify the function $f$ in this equation.
As in the inviscid case, the functional form of $f$ is
determined by requiring that the perpendicular hydro equations 
\eq{tranverse_eq} hold for the viscous case 
 
In order to find $f$ for the small perturbation, let us expand
\be
\label{mod_f}
f=T+ f'= T+T'+m
\ee
The first order terms in this expansion coincide with  inviscid fluid
since  viscous corrections vanish for the static homogeneous medium. 
If there were no viscous corrections,
$f'$ would coincide with the perturbed temperature $T'$ as in 
the ideal case. Thus the function  $m$ 
depends on the viscosity. 

Now consider a small perturbation of the hydrodynamic fields, 
$T'$ and  $u'^{\mu}$.
As in the previous section, we can split the velocity perturbation in a 
potential part that does not have vorticity $U^{\mu}$  and a rotational part
that carries the vorticity, $R^{\mu}$. We define 
now a potential related to the irrotational part of the velocity
perturbation using \eq{mod_f} as 
\be
\label{vp_def}
T'+m=u^{\mu}\partial_{\mu} \varphi  ~,\\ \nonumber
TU^{\mu}=h^{\mu \nu}\del_{\nu} \varphi ~,
\ee
This definition ensures that if $R^{\mu}$ is zero, the perturbed field
does not introduce vorticity. 

After this identification, the perturbation of the perpendicular
equation \eq{tranverse_eq} including shear viscosity can 
be written in terms of $R^{\mu}$ and $m$ only as
\be
\partial_{0} R_i+
\partial_{i} m = \frac{3}{4}\Gamma_s \nabla^2 R_i+
\Gamma_s \partial_i \nabla^2 \varphi ~,
\ee
where the right hand side comes from the viscous tensor which
has two independent contributions
coming from the irrotational and the rotational part of the perturbed
velocity field. Thus, identifying
\be
m = \Gamma_s \nabla^2 \varphi ~,
\ee
we absorb all the viscous effects coming from potential flow into
a modification of the function $f$.
After finding this modification due to the viscosity, we can use
the definition of the potential \eq{vp_def} into the (perturbed) entropy
continuity 
equation to find an equation for $\varphi$
\be
\label{dumped_wave}
\partial^2_{0} \varphi-c^2_s \nabla^2 \varphi- 
             \Gamma_s \partial_t \nabla^2 \varphi=0\,.
\ee
The equation for the irrotational part is
\be
\label{R_diff}
\partial_{0} R_i
 = \frac{3}{4}\Gamma_s \nabla^2 R_i \, .
\ee

\eq{dumped_wave} is the wave equation including dissipation 
coming from viscosity. This
equation for the potential coincides with the equation for the energy density
that one would find from our analysis in \cite{CST}, what is expected as the 
potential and the energy density in this case are related by a time derivative.
\eq{R_diff} is the diffusion equation for the rotational fluid
which coincides also with the equation for the diffuson in \cite{CST}.

\section{\label{EL} Energy and momentum  of small perturbations}

As argued in the introduction, an accurate matching of the
initial disturbance to the subsequent flow field is not
possible. We can only constrain global quantities such as the total energy 
and momentum deposited. In this section we study the total
energy and momentum of the modified fields in order to
learn about the possible mechanisms of interaction. 

To this end,  we consider a very high energy particle
that travels with finite velocity $v$ though an infinite homogeneous ideal baryon free 
fluid. We also assume
that the motion has taken place forever 
and is therefore static in the rest frame of the particle.
The analysis here parallels the discussion of 
non-relativistic supersonic flow past finite bodies \cite{landau}.

As shown in section \ref{SPD}, the equation for small potential perturbations of the potential
flow  can be readily
computed from the acoustic metric, leading to the usual wave equation
We will assume that the potential depends on the combination $x+vt$, with $v$
the velocity of the jet. With this constraint,  
the wave equation far from the disturbance reads
\be
\beta^2 \partial^2_{\kappa} \phi - \partial^2_{\perp} \phi=0 \, .
\ee
Here $\kappa=x+vt$, $\perp$ denotes the coordinates transverse to the jet
trajectory, and 
\be
\label{beta_def}
\beta^2=\left(\frac{v^2}{c^2}-1\right) \, .
\ee
In the region where this equation applies, we can write
general solution in the following form:
\be
\label{phi_fluid}
\phi=\frac{T}{2 \pi s} 
      \int^{x+vt-\beta \rho}_{-\infty} d\xi \frac{dF/dx(\xi)}{\sqrt{(x+vt-\xi)^2-\beta^2 \rho^2}},
\ee
The function $dF/dx(\xi)$ characterizes the source and will
be elucidated in section \ref{ad}.

From the discussion in section \ref{SPD} we know that 
the rotational part of the flow field $R^{\mu}$ (in the 
small perturbation regime) does not propagate and
its evolution is determined by the viscosity. Examining the kernel
in Eq.~\ref{kerg}, the rotational field far from the fluid is
\be
\label{Rgaus}
v^x=\frac{A}{\left( 2\pi \frac {3}{2 v} \Gamma_s (x+vt)\right)}
\exp \left\{- \frac{\rho^2}{\left( \frac {3}{2 v} \Gamma_s (x+vt)\right)} \right\} ~,
\ee
where the constant $A$ will be determined later\footnote{
The same expression is found in non relativistic fluids past bodies
where such field describes the non-turbulent 
wake behind the body \cite{landau}}. This field also
depends on $x+vt$.

In order to proceed further, it is beneficial to go to the rest frame
of the jet, where the unperturbed medium moves toward the
jet with velocity $v$.  In the 
jet rest frame,
the fields do not depend on the
proper time coordinate $\tau$ and depend only on the
longitudinal coordinate 
\footnote{Note that this conclusion depends on the fact that the
velocity of the jet does not change significantly on the
passage thought the medium which should be a good approximation 
for a high energy particle.}. 
$\chi=\gamma\left(x+vt\right)$.

\begin{figure}
\begin{center}
 \includegraphics[width=7cm]{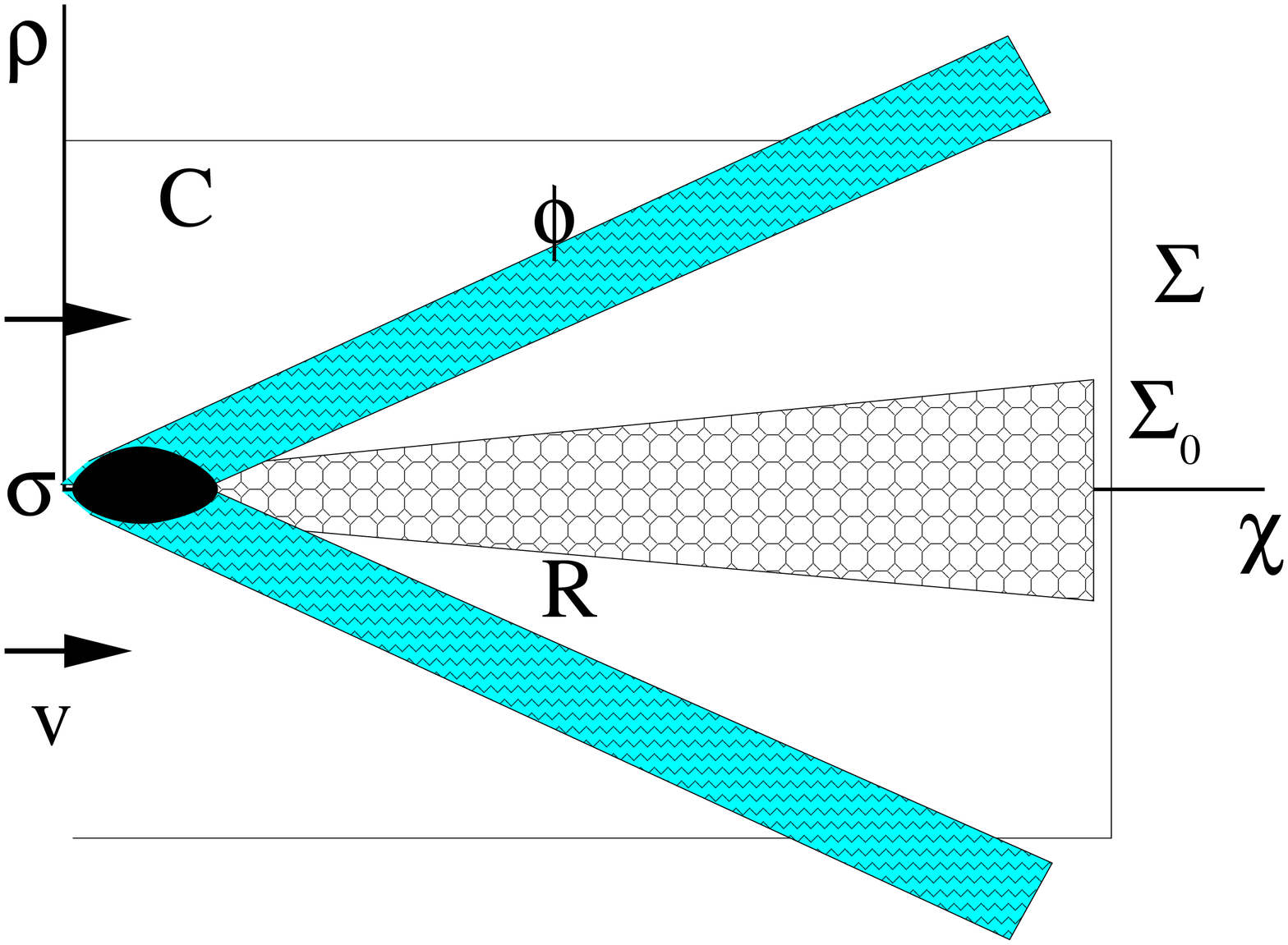}
 \caption{\label{cylinders}
Sketch of the flow picture in the fluid jet rest frame. 
The non-hydro core (solid region) serves as a source for
the hydrodynamic fields. The irrotational or potential component of 
the velocity field (wavy region) propagates out of the source
and leads to a small disturbance at large distances. The 
irrotational part(circled region) remains concentrated along the jet axis
within a transverse size that grows as $\sqrt{\chi}$.
}
\end{center}
 \end{figure}

The flow picture is sketched in figure \ref{cylinders}. The 
non hydro core co-moving with the jet has a typical transverse
size $\sigma$. The rotational flow remains concentrated along
$\chi$ to the right of the jet in a 
narrow wake 
 with a transverse extent that grows as $\sqrt{\Gamma_s \chi /v \gamma}$.
By contrast, the sonic disturbance
propagates and at sufficiently large distances from the
jet is  concentrated at the Mach angle within a typical
size given by $\sigma$. Note that for both types of excitations
the modified fields are always small sufficiently far away from the jet.
The fluid to the left of the jet cannot be modified
by the high energy particle and flows with a steady velocity, $v$. 

In this stationary situation, the energy and momentum loss can be
calculated by computing the momentum and energy flux out 
of a large cylindrical region surrounding the jet. 
Disregarding the left end  cap at $\chi = -\infty$ (where there is no 
modification), the surface integral includes the cylinder surface $C$,
and the right end cap, $\Sigma$.
The irrotational fluid only affects a small
subregion $\Sigma_0$ of the  right end cap, $\Sigma$ of 
area $\sim \Gamma_s \chi /v \gamma$.
 The potential flow contributes to
cylinder $C$ and to the rest of the end cap, $\Sigma-\Sigma_0$. Thus the 
energy lost by the jet is given by
\be
\label{dEcont}
-\frac{dE_j}{d\tau}=\int_C \left(T^{0\rho}\right)' +
\int_{\Sigma-\Sigma_0} \left(T^{0 \chi}\right)'+
\int_{\Sigma_0} \left(T^{0 \chi}\right)'
\ee
where $~'$ denotes the variation with respect of the unperturbed
field. 
As in the region $\Sigma_0$ the main modification of the fields is due 
to the rotational disturbances,
the two first terms in Eq. (\ref{dEcont}) are the energy
loss due to the potential flow, while the last is due to the rotational flow,
which we denote as 
 $-dE^R_j/d\tau$. Since far away from the jet
 the linear approximation is valid, we have
\be
\label{Tlin}
\left(T^{0\rho}\right)'=\left(sT\gamma^2 v^{\rho}\right)'=
                       T\gamma \left(s\gamma v^{\rho}\right)'+ 
                        s\gamma v^{\rho} \left(T\gamma \right)'~,
\nonumber \\
\left(T^{0\chi}\right)'=\left(sT\gamma^2 v^{\chi}\right)'=
                       T\gamma \left(s\gamma v^{\chi}\right)'+ 
                        s\gamma v^{\chi} \left(T\gamma \right)'~.
\ee
In both these expression we can identify the first term as the
modified entropy flux times a constant factor ($T\gamma$). 
The second term vanishes in the potential region
since the fluid is stationary and we have
$\left(T \gamma\right)'=\partial_{\tau} \varphi = 0$.
Next note the entropy produced per unit time is 
\be
\frac{dS_j}{d\tau}=\int_C \left( s\gamma v^{\rho}\right)'
                  +\int_{\Sigma-\Sigma_0} \left( s\gamma v^{\chi}\right)'
                  +\frac{dS^R_j}{d\tau}
\ee
where
$dS^R_j/d\tau$ is the entropy carried by the rotational flow in the $\Sigma_0$ 
region. The two quantities $dS^R_j/d\tau$ and  $-dE^R_j/d\tau$ are not
independent (what we show at the end of the section) but we prefer to
keep then explicitly in what is next to make our argument clearer. 
Putting these results together the energy loss is
\be
\label{GEloss}
-\frac{dE_j}{d\tau}=T\gamma \frac{dS_j}{d\tau} - 
\frac{dE^R_j}{d\tau}-T\gamma \frac{dS^R_j}{d\tau}
\ee
Thus, in the rest frame of the jet, the energy loss of the jet is related to
the entropy production minus corrections due to the rotational flow.

  Since the high energy particle remains almost on shell as it 
transverses the medium and looses energy, we conclude that the energy
and momentum loss rate are linked by the particle velocity $v$ as
$dE/dt=v dP/dt$. In the particle rest frame 
 we conclude that 
\be
\label{trivial}
\frac{dE_j}{d\tau} \approx 0~.
\ee
This relation immediately allows us to arrive to an important conclusion
about the interaction mechanism. 
From \eq{GEloss} and \eq{trivial}  we conclude that
if the interaction mechanism produces significant entropy, then
rotational flow is needed  
and the importance of the rotational flow with respect to the 
potential flow is related to the contribution of entropy 
production to the energy loss. Thus, we can distinguish two 
different types of interactions:

\begin{enumerate}
\item Isentropic interactions. The interactions of the jet and the fluid are
such that no entropy is produced in the process.
In this case the 
energy/momentum deposited can be calculated from the the far field sonic wave  
and it is quadratic in the perturbation.

\item Non isentropic interactions. The main mechanism of energy deposition
proceeds by transferring of heat into the fluid, creating new entropy. 
As a consequence, rotational flow is needed, which is concentrated in a 
narrow wake behind the fluid.
\end{enumerate}

Before continuing we perform a similar analysis for the momentum loss.
Integrating over a large cylinder as before, 
the momentum loss is 
\be
-\frac{dP^{\chi}_j}{d\tau}=\int_C \left(T^{\chi\rho}\right)' +
\int_{\Sigma-\Sigma_0} \left(T^{\chi \chi}\right)'+
\int_{\Sigma_0} \left(T^{\chi \chi}\right)'  \, .
\ee
Now, in the region of small perturbation  we find
\be
\label{dPlin}
\left(T^{\chi \chi}\right)'=\left(v^{\chi}T^{\chi 0}\right)'=
                               v\left(T^{\chi 0}\right)'+v^{\chi}T^{\chi 0}
\nonumber \\
\left(T^{\chi \rho}\right)'=\left(v^{\chi}T^{\rho 0}\right)'=
                               v\left(T^{\rho 0}\right)'+v^{\chi}T^{\rho 0}
\ee
where we identified a term proportional to the energy loss (due to the 
potential part of the flow).
Denoting, $d\left(P^{\chi}_j\right)^R/d\tau$ 
the momentum loss due to 
the rotational flow (which contains the sum of the momentum flux through
the surface $\Sigma_0$ and  $v{dE^R_j}/{d\tau}$),
the total momentum loss is given by
\be
\label{GPloss}
-\frac{dP^{\chi}_j}{d\tau}=-v\frac{dE_j}{d\tau}
                            +\int_C Ts \gamma^2 v^{\chi}v^{\rho}
                            +\int_{\Sigma-\Sigma_0}  v^{\chi}T^{0\chi}
                            -\frac{d\left(P^{\chi}_j\right)^R}{d\tau}
\ee
where we separated once again the contributions for the rotational and
irrotational part of the flow as in \eq{GEloss}. This
equation will be useful  in the next section. 

To complete this section, let us relate the energy and momentum loss and 
the entropy rate in the wake to the
amplitude of the rotatinal field. Boosting Eq. (\ref{Rgaus}) to the jet rest 
frame we find $v^{\chi}=v^x/\gamma^2$, and $v^{\chi}$ is a Gaussian in the
transverse coordinate $\rho$ with width depending on $\chi$. 
From Eq. (\ref{dPlin})
\be
\label{Pwake}
-\frac{d\left(P^{\chi}_j\right)^R}{d\tau} =- T s v \gamma^2 \int_{\Sigma_0} 
v^{\chi}=   
 -sT v A
\ee                           
In the same way, by expanding Eq.(\ref{Tlin}) and using the fact that
$dE_j/d\tau=0$ we obtain
\be
\label{Swake}
\frac{1}{\gamma} \frac{dS_j}{d\tau}=\frac{1}{\gamma} \frac{dS^R_j}{d\tau}
+\frac{1}{T\gamma^2}\frac{dE^R_j}{d\tau}=
v^2 s A
~.
\ee
%
%

\section{\label{ad} Isentropic Interactions}

As argued in the previous section, a purely potential excitation
of the fluid by a high energy particle implies that no significant
entropy can be produced in the interaction. In this case, the
perturbed flow field in the far region is given 
by \eq{phi_fluid} and 
depends on a unknown function
$dF/dx(\xi)$ that characterizes the source.
To clarify the physical meaning of this source, let $dF/dx(0)$ mark the 
smooth beginning of the source which moves with the jet velocity.
Further assume that the typical scale of variation of this function $\lambda$ is much larger that the transverse distance at which \eq{phi_fluid} is applicable, {\it i.e.} the source is elongated. 
In this case, for $\rho \ll \lambda$ and far 
away from the tip of the source, we can approximate \eq{phi_fluid} with
\be
\label{laplaze_sol}
\phi=-\frac{T}{2 \pi s} dF/dx(x+vt) \ln \left(\frac{\beta \rho}{2(x+vt)}\right),
\ee
Thus, for transverse distances short compared to the length of the jet,
the velocity of the perturbation $v^{\rho}$ decreases as $1/\rho$, leading to
finite radial flux of entropy out of a cylinder close to the jet. 
With this normalization $dF/dx(\xi)$ is the entropy flux per unit length at a position
$(\xi)$ with respect to the beginning of the jet.
This is completely analogous with the non-relativistic case, where
the coefficient is related with the mass flux 
\footnote{
In the nonrelativistic case,  matter is pushed out   of the
cylinder by the profile of the solid body that moves in the medium.
In our case there is
 no solid wall 
which prevents  the liquid from reaching center of the jet.
However, fluid is pushed out this region by  
interaction of the medium with the high $p_T$ particle. We will not speculate further about these microscopic details.
}
\cite{landau}.

As already stated, this analysis is identical to the non-relativistic case
of flow past a solid body. As in that case, we see that there is the formation
of sound waves, or vorticity free excitation that fall slowly far form 
the body. This slow fall means that the waves carry energy and momentum out
of the body, and we can compute the momentum drag  from the far field solution where the  linearized
approximation is justified. This loss by sonic emission is called sonic
drag.

In order to compute the momentum losses
 and to use standard techniques from text books \cite{landau}, it is
beneficial to go to the rest frame 
 of the jet by performing a Lorentz
transformation, as done in the previous section. We will
see that in the final expression for the energy and momentum loss
all the $\gamma$ factors drop out, so we can use the expressions obtained 
for the case of finite mass for massless particles.

As $\phi$ is a scalar, transforming to the jet rest frame means simply 
the transformation of the coordinates. Thus, by performing the change of 
variables $\xi \rightarrow \xi \gamma$ and defining 
$A(\xi)=dF/dx(\xi/\gamma)$, we obtain
\be
\label{phi_plane}
\phi=\frac{T}{2 \pi s}\int^{\chi-\bar{\beta} \rho}_{-\infty}
     \frac{A(\xi)}{\sqrt{(\chi-\xi)^2-\bar{\beta}^2 \rho^2}} d\xi ~,
\ee
with $\chi$ the longitudinal coordinate in the jet frame and 
$\bar{\beta}=\beta \gamma$. As we note in the previous section,
there is no dependence of the 
proper time $\tau$ (stationary situation).

We now express the energy and momentum loss expressions 
\eq{GEloss} and  \eq{GPloss} for the particular case where
there is neither entropy production nor rotational flow.
In this case the energy loss in the jet rest frame vanishes 
identically. By using the definition of the perturbed potential,
$T u'_{\mu}=\partial_{\mu}\varphi$ and noting that 
the perturbed velocity field $v^{\chi}$ vanishes as $ \chi \rightarrow \infty$ 
  we conclude that 
\be
\label{loss}
\frac{dE_j}{d\tau}=0~, \\ \nonumber
\frac{dP^{\chi}_j}{d\tau}=  -\int_C \frac{s}{T}\partial_{\chi} \phi \partial_{\rho} \phi ~, 
\ee
where $\int_C$ denotes the integral around a cylindrical surface
far from the fluid.
\eq{loss} gives
the correct expression of the energy momentum loss by the high $p_T$
 particle if
there is neither entropy nor vorticity production.

This is exactly the same expression that is obtained in the nonrelativistic 
case, but replacing the corresponding quantities as already explained. We can
then just refer to standard textbooks \cite{landau} to write the final 
value of the integral.
\be
\label{div_p}
\frac{dP^x_j}{d\tau}=-\frac{1}{4\pi}\frac{T}{s} \int^L_{-L} d\xi_1d\xi_2
                   \dot{A}(\xi_1) \dot{A}(\xi_2)
		   \left(\ln|\xi_1-\xi_2| -\ln(4L) \right)~,
\ee
where L is the longitudinal size of a long cylinder ($L\gg\lambda$)
and $\dot{A}$ denotes the derivative.
 The last term
in \eq{div_p} diverges in the limit $L\rightarrow \infty$ unless the 
flux of entropy out of the cylinder vanish far away from the jet.

We now perform a Lorentz transformation to the fluid frame. We start by noting
that according to our definition of $A(\xi)=dF/dx(\xi/\gamma)$, we can 
change the variable of integration in \eq{div_p} and in the limit 
$L\rightarrow \infty$ the expression is independent of $\gamma$.
The transformation $dt/d\tau=\gamma$ also cancels the $\gamma$ factor of the
Lorentz transformation leading to an expression for the energy and momentum
loss that is applicable when $v \rightarrow c$
\be
\label{dedt}
\frac{dE}{dt}=-v\frac{dP^x}{dt}= -v \frac {dP^x_j}{d\tau}~,\nonumber \\ 
\frac {dP^x_j}{d\tau}=-\frac{1}{4\pi}\frac{T}{s} \int^{\infty}_{-\infty}
                      d\xi_1d\xi_2
                   \frac{d}{d\xi}{dF/dx}(\xi_1)\frac{d}{d\xi}{dF/dx}(\xi_2)
		   \ln|\xi_1-\xi_2|~.
\ee

Note that, as a consequence of the first equation in \eq{dedt},
the wave drag is such that the
particle remains on shell. The relative sign between the energy and 
momentum loss is due to the fact
that, according to our conventions, the jet is moving in the $-x$ direction.
As the particle remains on shell, the isentropic assumption is a fully
consistent way to describe the interaction of the jet and the medium.
Note also that we assume that 
the speed of the propagation did not change, what should not be important for 
particles moving with velocity close to $c$.

\section{\label{nad}Production of Entropy}
We discuss now the case where the significant entropy is produced in the
interaction of the jet with the medium. 
In this case the disturbance 
remains close
to the interaction region in a narrow wake.
In the wake, hydrodynamics and 
linearized hydrodynamics are not justified until
the viscosity dissipates the strength of this
disturbance.
At the end of the section we will also briefly discuss
the spectrum of particles produced for such excitations. 

In the language of infinitesimal disturbances of section \ref{LH}
this case can be accounted for by 
initial condition  such that $\epsilon_0(t_0,{\bf x})$ has a finite integral 
over space. In the linearized region the total field is found
by summing these infinitesimal disturbances. 
In this setting, it is evident that the energy loss is fist order in
the perturbation as, to this order, it coincides with the integral
of the function $\epsilon_0$

As seen in the section \ref{EL}, the requirement that 
the particle remains almost on shell on its propagation
implies that the production 
of entropy should be accompanied by vortex flow. 
In the language of section \ref{LH} it corresponds to a 
function ${\bf g_0}(t_0,{\bf x})$ which cannot be expressed as a 
gradient $\left({\bf \nabla} \times {\bf v} \neq 0 \right)$
and with finite integral of the ${\bf g^z_0}$ component.
Such a case was studied in \cite{CST} with the following particular
expressions for the initial conditions:
\be
\epsilon_0(t_0,x)= g^z_0(t_0,x)=\frac{dE}{dx} 
        \frac{e^{-\left({\bf x}-{\bf r_j}(t_0)\right)^2/ 2 \sigma^2} }
                                      {\left(2 \pi \sigma^2 \right)^3}~,
\ee
where ${\bf r_j}(t_0)=(t_0,-t_0,0,0)$ is the jet position. In
this case, the energy and momentum loss is $dE/dx$, and it is 
first order in the perturbation. Let us note that, 
as after setting the value of the fields at $t_0$ we only solve
the field equations for $t>t_0$, the final solution obtained as the 
addition of all the disturbances at different times $t_0$ 
takes the form
\be
T'^{0 \nu}=\int d{t_0} \theta(t-t_0) \delta T^{0 \nu}(t_0) ~,
\ee
where $\delta T^{0 \nu}$ is a solution of the linearized field
equations with initial conditions specified at $t_0$ (section \ref{LH}). Thus,
the differential equation that our solution satisfies is, in 
fact, 
\be
\label{source}
\partial_{\nu} T^{\mu \nu}=J^{\mu}~,
\ee
where $J$ depends on the fields at each $t_0$ which 
in the particular case used is
\be
J^{\mu}= \frac{dE}{dx} 
        \frac{e^{-\left({\bf x}-{\bf r_j}(t)\right)^2/ 2 \sigma^2} }
                                      {\left(2 \pi \sigma^2 \right)^3}
         \cdot \left( 1,-1,0,0\right)~.
\ee 

Let us remark here that the source \eq{source} used in
\cite{CST} is the same used in \cite{Heinz} to solve 
hydrodynamic equations in the nonlinearregime in 
an expanding, boost invariant, ideal, baryon free fluid
\footnote{ the actual calculation in \cite{Heinz} integrates several
of these sources along the space rapidity $\eta$ in order to keep boost 
invariance but the conclusions should not change}.
Thus, our solution is the linearized version of this problem
for a homogeneous medium.

After reexpresing the fields in this way, we want to connect the 
description of the fluid in terms of section \ref{pot} and show
that excitations where entropy is produced the modification of
the fields are mainly concentrated in the near region.
This is clearly seen from the rotational component (excited by $g^z_{t_0}$).
 As demonstrated in section
\ref{SPD}, for a static medium the vorticity of the fluid does not
propagate, and the disturbance remains close to the deposition 
region, {\it i.e.} in a region of typical size $\sigma$ 
the source size, and thus
far from  the source size the fluid is potential.
 This was also found in \cite{CST} by
solving the equations of motions including viscosity.

Thus, if the interaction jet liquid is such that significant total 
entropy is produced, then the modification of the hydro fields
are mainly produced in the near region (the `non-hydrodynamic core)
and, as argued in the 
introduction, they should not be described  in terms of ideal hydrodynamics;
one instead would need to use dissipative hydro in order to gain some 
understanding from this region.

To conclude this section let us briefly discuss the particle production 
for medium excitations that produced entropy, which will be 
extensively done in  the next section for the entropy free case.
As found in \cite{CST}, the excitation of the diffusion mode (vorticity)
leads to matter moving preferentially in the direction of the jet.
This flow is due to the region close (distances of the order of $\sigma$)
to the jet path and leads to a complete shadowing of any Mach cone signal
in the dihadron azimuthal distributions, resulting in correlations only 
at $\Delta \phi = \pi$. Similar conclusions have been reached 
in \cite{Heinz}, confirming our result for the linearized static theory in the 
non linear expanding regime. Thus, this kind of excitations, if 
present, leads to a fill up of the Mach cone in the two particle correlations
associated to the jet.  For more details we refer the reader to our
short paper \cite{CST} and to \cite{Heinz}. Let us remark once again
that our calculated spectrum in \cite{CST} is dominated by the near 
region, where the non hydrodynamic core is, and thus, the description in terms
of hydrodynamics is not justified.

\section{\label{NSP} Spectra and correlations associated 
with the jet 
}
We now calculate the spectrum of particles associated with 
a jet through the induced disturbance of the medium. We will consider,
as in the previous discussion, a jet propagating in a static
homogeneous baryon free fluid. The jet is created at some time $t_0=0$
and is absorbed due to the energy loss  at some later time $t_j$.
We take the life time of the jet $t_j$ as a parameter,
which means that, for a fixed value of the energy loss, we are varying
the energy of the jet.
 In a realistic simulation, this time would be given
by the energy loss and the spectrum of produced jets.
 Finally, the system freezes out at a time 
$t_f$
where we calculate the spectrum though the Cooper Fry 
prescription \cite{cf}.
We will always consider $t_j<t_f$ so that the final spectrum is not dominated
by our arbitrary source.

We will only study the case where there is no entropy production, 
and consequently the energy loss is second order.
 We will consider the
superposition of infinitesimal disturbances described in section
\ref{LH} and demand that no entropy is produced. 
This can be achieved by  considering infinitesimal
displacements such that the modification of the fields at a time $t_0$ 
in a time interval $dt_0$ is
\be
\epsilon_{dt_0}=0~, \\ \nonumber
 g^i_{dt_0}=\frac {C}{\sigma^4}\partial_i e^{-r^2/2\sigma^2}~.
\ee
The source
size $\sigma$ is the characteristic distance from the jet 
such that the 
linearized approximation is valid. 
The amplitude of the disturbance is  
 $C/\sigma^4$ with 
$C$ a dimensionless constant. The factor
$1/\sigma^4$ is introduced {\it ad hoc} such that the disturbances
have the correct dimensions keeping $C$ dimensionless. 

As explained in the introduction, the process by which the energy
of the jet is thermalized is highly dissipative and we cannot describe it.
Thus, we cannot fix from first principles 
the two parameters of the initial perturbation $C$ and $\sigma$ 
 unless a good matching
mechanism is found. Thus we will use the energy and momentum loss
to relate their values. 
Following appendix \ref{matching} we can relate
this disturbance with a far field solution  in the form \eq{phi_fluid}.
After appendix \ref{matching}, we identify the function $dF/dx$   
in term of the parameters $C$ and $\sigma$ as
\be
\frac{dF}{dx}(\xi)=-\frac{2\pi}{T}\frac{v^2}{c^3_s}\frac{C}{\bar{\sigma}^2}
                   \del_{\xi}e^{-\xi^2/2\bar{\sigma}^2}~,
\ee
where we have introduced for simplicity $\bar{\sigma}=v\sigma/c_s$. 
With this identification we can evaluate the energy loss using \eq{dedt}
leading to
\be
\label{Elinv}
\frac{dE}{dt}\approx \frac{\pi^2}{v}\frac{1}{sT}\frac{C^2}{\sigma^6}~.
\ee

As expected, the energy loss is quadratic in the amplitude of the 
disturbance, $C$. The strong dependence on the typical size $\sigma$
is partly artificial and comes  from our choice of using
$\sigma$ as the dimensionfull parameter while keeping  $C$ dimensionless.

We will also consider that the static medium has a non vanishing shear
viscosity that leads to a reduction of the amplitude of the sound waves
as it propagates out of the disturbance. As this attenuation of the wave
can be understood \cite{landau} as a result of the dissipation of 
the mechanical energy (given by \eq{Elinv}), we will use the previous
formula for the energy/momentum loss also in the viscous case.

Let us summarize here the parameters that we have in our calculation. 
The source has two parameters, the amplitude $C$ and the source size $\sigma$.
This two can be constrained by the value of the energy loss. The timings
are characterized by the jet lifetime $t_j$ and the observation time 
$t_f$. In a realistic simulation, the lifetime is determined from the
jet energy and energy loss; the freeze out, time from the hydrodynamical
simulations at RHIC. Finally, the medium is characterized by the
speed of sound $c_s$, the sound attenuation length $\Gamma_s$ and the
temperature of the static medium. We will fix the speed of sound
to be $c_s=1/\sqrt{3}$ and we will not consider its change here in spite
of our findings in \cite{CS} that showed that those changes
are very important for the final observation and the position of the peaks.
The value of $\Gamma_s$ is taken close to its minimal bound
$\Gamma_s=1/4\pi T$ \cite{visc_bound}. Finally, everything is calculated
 in units of the
temperature, that should be of the order of the critical temperature. 
From now on we will assume that the jet moves at the speed of light
$v=1$. 
Let us finally give a numerical value by substituting 
the QGP equation of state as measured by lattice calculation \cite{karsch}
($e\approx 12 T^4$)
\be
\frac{dE}{dt}\approx 0.63\frac{C^2}{\sigma^8} \sigma^2 \frac{1}{T^4} ~.
\ee

Once we have listed all the parameters that enter in our calculation
we can calculate the spectrum of particles induced by the jet. As the 
medium is static, we use as the Cooper Fry prescription for equal time
freeze out surface as in \cite{CST}
\be
\label{mspectrum}
\frac{dN}{d^3p}=\int_V \frac{d^3V}{2\pi^3} e^{-\frac{E}{T}+\delta}~,
\ee
where V is the volume of the fireball, $T$ is the temperature at 
freeze out and $\delta$ the perturbation of the fluid. For $\delta=0$ 
 we obtain the unperturbed spectrum from the static medium.
When a jet goes through the fluid, $\delta$ is related to the modified
fields as:
\be
\delta= \frac{E}{T} \frac{\delta T}{T} + \frac{\vec{p}\vec{v}}{T} ~.
\ee

From these expression we expect non trivial features of
the $p_T$ dependence of the correlation function. 
As discussed in \cite{CST}, we can distinguish two different regimes
in the particle production:

\begin{itemize}
\item[1.] Low energy particles $E\sim T$. In this region 
we can expand the exponential in \eq{mspectrum}
and express the spectrum in terms of the energy and momentum deposited 
\be
\label{lowspc}
\frac{dN}{d^3p}=\frac{e^{-\frac{E}{T}}}{(2\pi)^3}\left(V+\frac{E}{T}
                \frac{E_{dep}}{w} 
                +\frac{\vec{P}}{T}
                  \frac{\vec{P}_{dep}}{w}\right)
~.
\ee          
The term proportional to the fireball volume $V$ corresponds to
the uncorrelated contribution that is subtracted from the dihadron 
correlation function. 
The other two terms are due to the disturbances. As seen in \eq{lowspc},
the soft particles are insensitive to the particular shape
                of the flow field and the
 angular dependence  is just a cosine
of the relative angle between the observed particle and the jet. Thus,
linearized hydrodynamics  is unable to produce peaks 
for very soft particles.

\item[2.]  High energy particles $E\gg T$. In this region the large parameter
$E/T$ compensates the small flow velocity. The integral is then
 dominated by the 
maximum of the exponent. Thus, the final spectrum reflects the shape of the
sonic disturbance and  only points of maximum modification of the hydrodynamic
fields contribute to the integral. 
This fact is also
responsible for the absence of the conical flow
when significant entropy is produced. In this 
case  the region of maximum modification
is close to the jet and the spectrum is dominated by the near
field region where the applicability of the hydrodynamic approach is 
uncertain. 
\end{itemize}

Finally, we calculate the spectrum of 
correlated particles by subtracting the 
unperturbed spectrum ($\delta=0$) from the jet induced 
spectrum ($\delta \neq 0$) 
\be
\frac{dN}{dyd\phi}=\frac{dN}{dyd\phi}\Bigg |_{\delta \neq 0}
                  -\frac{dN}{dyd\phi}\Bigg |_{\delta = 0} \, .
\ee

\subsection{Comparison with experimental data}

\begin{figure}[t]
\begin{center}
\includegraphics[width=6cm,height=10cm]{dNdydphipt_SL.epsi}
\includegraphics[width=7cm]{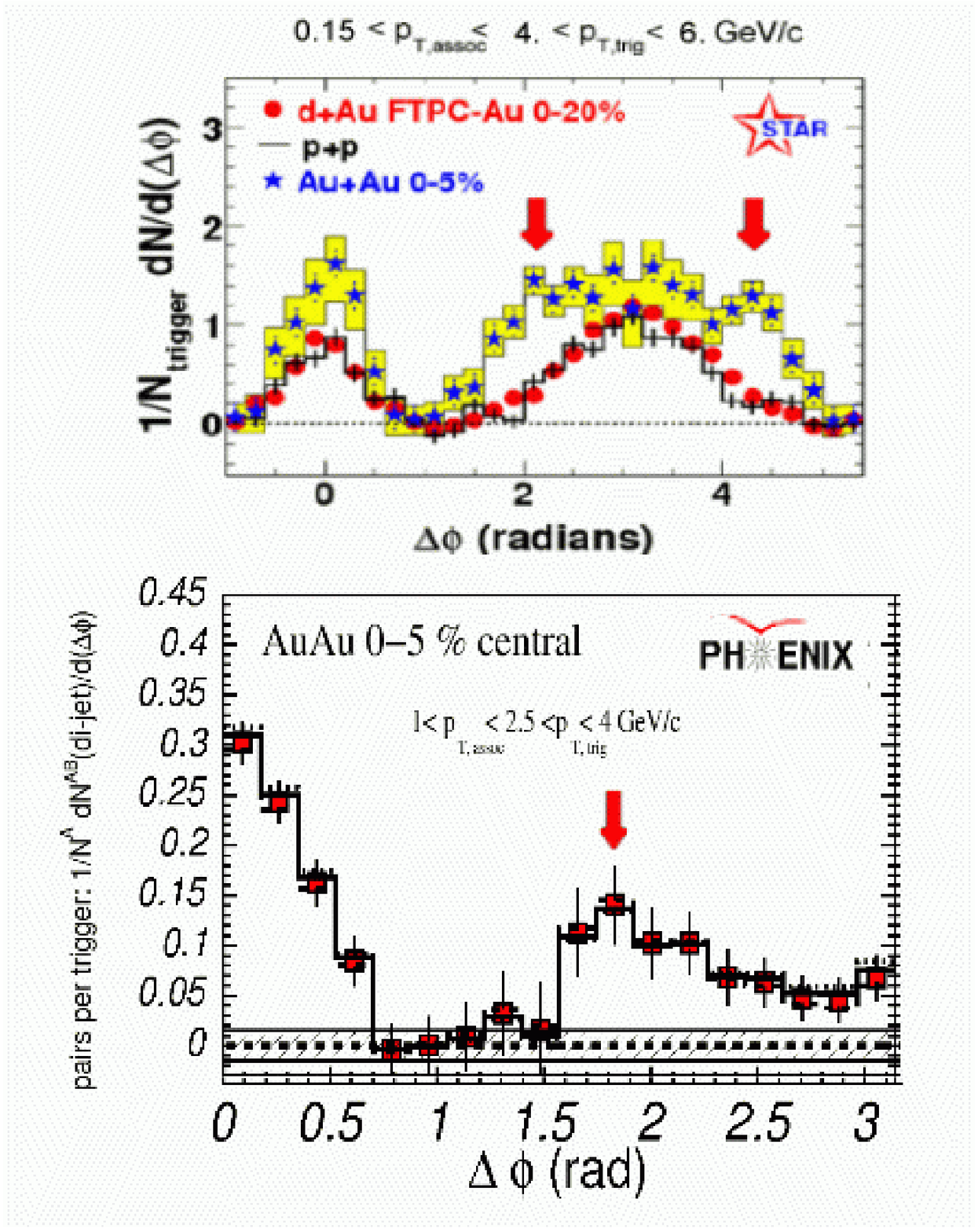}
\end{center}
\caption{
\label{corptdep}
Left: Associate yield dependence on associate $p_T$ 
for fixed source size $\sigma=0.75/T$
 , viscosity $\Gamma_s=0.1/T$, $t_j=8/T$, $t_f=10/T$,
and energy loss, $dE/dx=10 T^2$ (top) and $dE/dx=63 T^2$ 
(bottom). The label values for $dE/dx$ correspond to 
$T=200~\mbox{MeV}$.
 The three curves are for  
 $1 T<p_{t}<5 T$ (solid),
 $5 T<p_{t}<10 T$ (dotted),
($3\times$) $10 T<p_{t}<15 T$ (dashed),
($10\times$) $15 T<p_{t}<20 T$ (dashed-dotted).
(in the upper panel all the curves are rescaled further up
by a factor 10). No large angle correlation is observed
for $dE/dx=10 T^2$. For $dE/dx=63 T^2$
the position of the peak shifts toward $\pi$
for lower $p_T$.
Right: Experimental dihadron azimuthal distributions from
STAR (top) \cite{star_peaks} and 
PHENIX (bottom) \cite{phenix_peaks}
}
\end{figure} 

From the previous discussion of the spectrum we expect that for 
sufficiently large energy particles, the azimuthal distribution
of particles associated with the jet should reflect the modification
of the flow fields due to the particle passage. As argued in \cite{CST}
this translates into the appearance of large angle correlations in
the azimuthal dihadron distribution associated with the jet. Such large 
angle correlations have been observed by both the PHENIX \cite{phenix_peaks} 
and STAR \cite{star_peaks}
collaborations.

In this section we will fix a set of values for our parameters
 that give 
similar dihadron correlation than those in \cite{phenix_peaks} 
with no other justification as to reproduce qualitative features of 
observed on experiment. Those values are $\sigma=0.75/T$,
 $\Gamma_s=0.1/T$, $dE/dx=6.3 T^2$, $t_j=8/T$ and $t_f=10/T$.
The corresponding correlations functions are shown in the
lower left panel of figure \ref{corptdep}. 
Let us note that for a typical temperature of the plasma
of the order of $T\approx 200 ~\mbox{MeV}$, the required value 
of the energy loss is very large $dE/dx=12.6 ~\mbox{GeV/fm}$.
If we however would use a more realistic value of the energy loss
$dE/dx =2~\mbox{GeV/fm}$ we are not able to generate any large angle
correlation or magnitude of the correlation for the intervals
of momentum considered,
 as shown in the upper left panel of figure \ref{corptdep}.
\footnote{
Let us note however that, according to \cite{EHSW} RHIC data seem to
support values of $\hat{q}$ that are up to 4 times larger than those
of the peturbative estimates from \cite{Baier}, which lead to energy
losses that are much closer
to our required value.}

Even though the necessary value of the energy loss within our static
approximation is very large, we will try to argue in the next section,
 where we discuss the dependence of the
calculated spectrum on the different parameters, that 
expansion effects may help to reduce the necessary energy loss.
That is why we want to point out here some qualitative agreement
of our calculation (with large energy loss) with the experimental
data. This is the $p_T-$dependence of the correlation.

In the lower left panel of 
 figure \ref{corptdep} ($dE/dx=12.6~\mbox{GeV/fm}$)
 we show the $p_T$ dependence of the 
azimuthal dihadron correlation for different $p_T$ windows.
The  four curves correspond to  
$p_T$ intervals  (in units of temperature) of 1-5,5-10,10-15,15-20 for
solid, dotted, dashed and dashed dotted respectively. Note that we have
rescaled up the two highest $p_T$ bins by factors 3 and 10 respectively. 
This figure shows clearly the behavior of the correlation function. As
claimed, at 
low $p_T$, when the ratio $p_T/T$ is small (solid line),
 the formation of conical flow in 
the Mach direction is not reflected in particle spectra; we obtain, in 
fact, that  
particles are produced mainly in the direction opposite to the jet.
As $p_T$ increases, the correlation develops, moving the peak to the left
toward the Mach angle and making the width of the peak at this angle
smaller. In the upper left panel, for smaller values of $dE/dx$ (where
all the curves are further rescale up by 10),
the correlation always peaks at $\Delta \phi= \pi$ for the intervals 
of $p_T$ considered. This is because the fields in this case are too 
week. 

In figure \ref{corptdep} (right panel)
 we show the dihadron correlation function
as presented by STAR (top) \cite{star_peaks} and PHENIX (boton) 
\cite{phenix_peaks}. Even though our calculation cannot be directly
compared to the experimental situation, as we have not included the
effect of the expansion \cite{S_w,Renk} nor the variable speed of 
sound \cite{CS}, that may be very important for the final observation
of the effect, some qualitative features seem to be reproduced. In fact,
the correlation function shown by STAR, seems to be a broad peak with a 
small peak at a finite angle away form $\pi$ (marked by the red arrows).
This could be  explained from our discussion of the $p_T$ dependence
as in the experimental conditions of STAR, where the associated $p_T$
is dominated by soft particles ($0.15<p_T<4\,\mbox{GeV/c}$), and thus,
according to the previous discussion, the strength of the correlation 
should be small, as is due to the high-pt tail of the associated particles.
On the contrary, in the case of PHENIX, where the associated particles
are harder ($1<p_T<2.5\,\mbox{GeV/c}$), the correlation function shows a well
defined peak off $\pi$. What is more, for the particular values of the
parameters chosen, the amplitude of our calculated correlation function
corresponds roughly with the experimental magnitude observed.

Let us remark here that the $p_T$ dependence of the correlations coming
from the conical flow, which seems to be in agreement with the experimental
situation, is completely different from those radiative based mechanism
to explain large angle correlation \cite{Dremin, MWK, Vitev}. These model
predict that the correlations should shift toward $\Delta \phi=\pi$
as the $p_T$ of the associated particle increases, which is the opposite
to what we find.

The position of the peak in our correlation function is at 
$\Delta \phi \approx \pi - \arccos\left(1/\sqrt{3}\right) = 2.2 ~\mbox{rad}$. 
 and
 is set by the value of the speed of 
sound in the medium, which in our case is the ideal
QGP value of $c^2_s=1/3$. This is of course not the case
in the experimental condition at RHIC, where as the medium expands
and cools, the speed of sound changes from the previous QGP value
to $c^2_s\approx 0$ in the mixed phase and $c^2_h\approx 0.2$ in the
hadron gas. From hydrodynamical simulations of Au-Au collisions at RHIC
leads to similar (proper) time extension of the three previously mentioned
phases of approximately $\tau \sim 4-5 fm$. As argued in \cite{CST} 
the Mach angle depends only on the ratio of distance traveled by the 
disturbances to the one traveled by the jet; thus the position of the
peak is modified due to the expansion and can be estimated from the
average speed of sound along the evolution of the medium
\be 
\label{eqn_Mach}
\cos\theta_M=\frac{1}{ c\tau}\int_0^\tau c_s dt\approx 0.333 ~
\Longrightarrow \Delta \phi \approx \pi - 1.2 ~\mbox{rad}.~,
\ee 
This value agrees nicely with the experimental value shown in
the experimental correlation function as seen in figure \ref{corptdep} b)

\begin{figure}[t]
\begin{center}
\includegraphics[width=10cm,angle=-90]{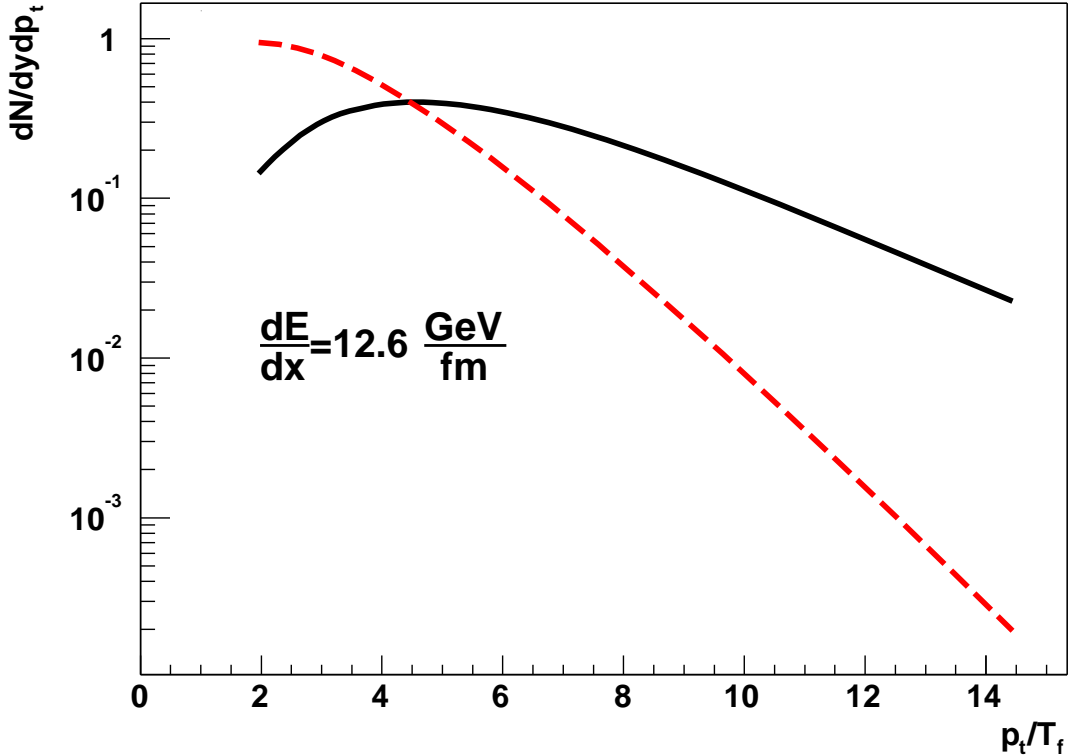}
\end{center}
\caption{
\label{awayS}
Spectrum of associated particles in the away side $\Delta \phi >1$
for  $\sigma=0.75/T$, $dE/dx=142 T^2$, $\Gamma_s=0.1/T$ (solid).
Spectrum of uncorrelated particles (rescaled down by a factor 100).
The associated yield is much harder than the inclusive due to the 
boosted liquid induced by the jet.
}
\end{figure} 

Finally, in figure \ref{awayS} we show the spectrum of correlated particles 
in the away side jet.
We defined the away side as particles as
$\Delta \phi >1$ as in \cite{star_peaks}. We compare it with the spectrum
obtained for uncorrelated particles, as if no jet were produced. In order to
be able to show both in the same plot, 
we rescale down this last spectrum (that we will call
inclusive) by a factor hundred. We observe that the correlated spectrum
is harder than the inclusive one. This is due to the fact that the liquid 
is boosted due to the non zero velocity fields in the Mach direction. 
The experimental spectrum shown in \cite{star_peaks} also seems slightly 
harder than the inclusive one, however the effect is by no means so 
prominent. On the other hand our inclusive spectrum is much steeper than
those in \cite{star_peaks} as our background fluid lacks radial flow. Thus,
even though there is no quantitative agreement between our calculation and
STAR data in this issue, it could be due to the static approximation that 
we took for the liquid, that is definitely not true.

\subsection{Dependence on the parameters}

\begin{figure}[t]
\begin{center}
\includegraphics[width=10cm,angle=-90]{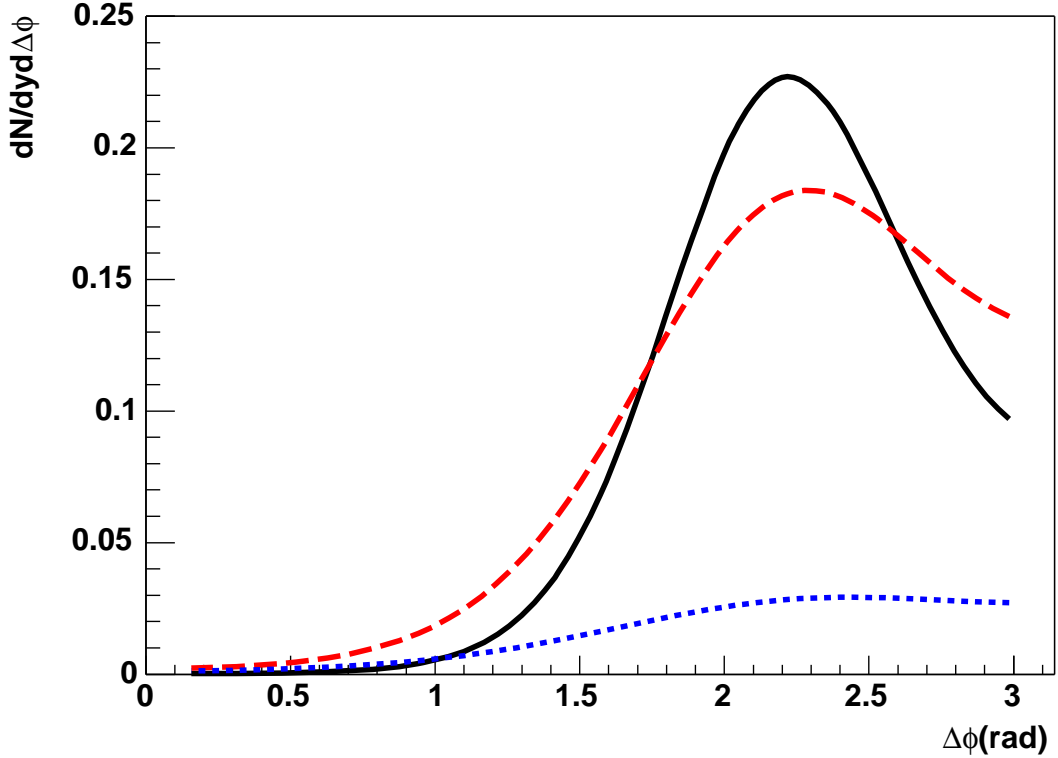}
\end{center}
\caption{
\label{dmdep}
Associate yield dependence on energy loss for fixed source size 
$\sigma=0.75/T$,
$\Gamma_s=0.1/T$, $t_j=8/T$, $t_f=10/T$  and $10 T<p_{t}<20 T$.
The three curves are for $(1/20 \times) dE/dx=126 T^2$, $dE/dx=63 T^2$,
$dE/dx=35.5 T^2$ for solid, dotted and dashed respectively. }
\end{figure}

In the previous subsection we saw that in order to produce similar correlation
functions as the experimental ones we need too large values for the energy
loss. In this section we want to show that our calculation is very sensitive
to the parameters chosen.

In figure \ref{dmdep} we study the dependence on the energy loss (or the source
amplitude) for the rest of the parameters fixed. We vary the energy loss
by a factor of two from the central value of $dE/dx=63 T^2$
({\it i.e.} 25.2, 12.6, 6.3 $\mbox{GeV/fm}$ for $T=200~\mbox{MeV}$). We observe
an extremely strong dependence on the amplitude (note that 
the solid curved is rescaled down by a factor 20). This is easily 
understood as from \eq{mspectrum} we see that the amplitude enters in 
the exponent and thus small changes lead to large variations on the final
observation. This strong dependence is the basis for our claim that 
expansion effects may be very important for the final observation of the 
effect on experimental conditions as it can reduce the needed values of 
$dE/dx$.

In \cite{CS} two of us found, by studying a simple model for a dynamic medium
(that of a static fluid in an expanding universe), that
the expansion of the liquid and dropping of the speed of sound leads
to an enhancement of the ratio $v/T$. Estimates made for  RHIC tell us
that for disturbances originated at early times, such enhancement can be
as big as a factor 3. What this means for the spectrum is that
at the time of freeze out, the amplitude of the disturbance can be 
up to a factor 3 smaller than in the
static case to reproduce the same correlation
(as the smaller amplitude is compensated by the dropping density and speed of 
sound of the background medium). Such a smaller value of the amplitude
enters quadratically into the energy density what means that the energy loss
can be up to a factor 9 smaller than those showed in the static case.

\begin{figure}[t]
\begin{center}
\includegraphics[width=10cm,angle=-90]{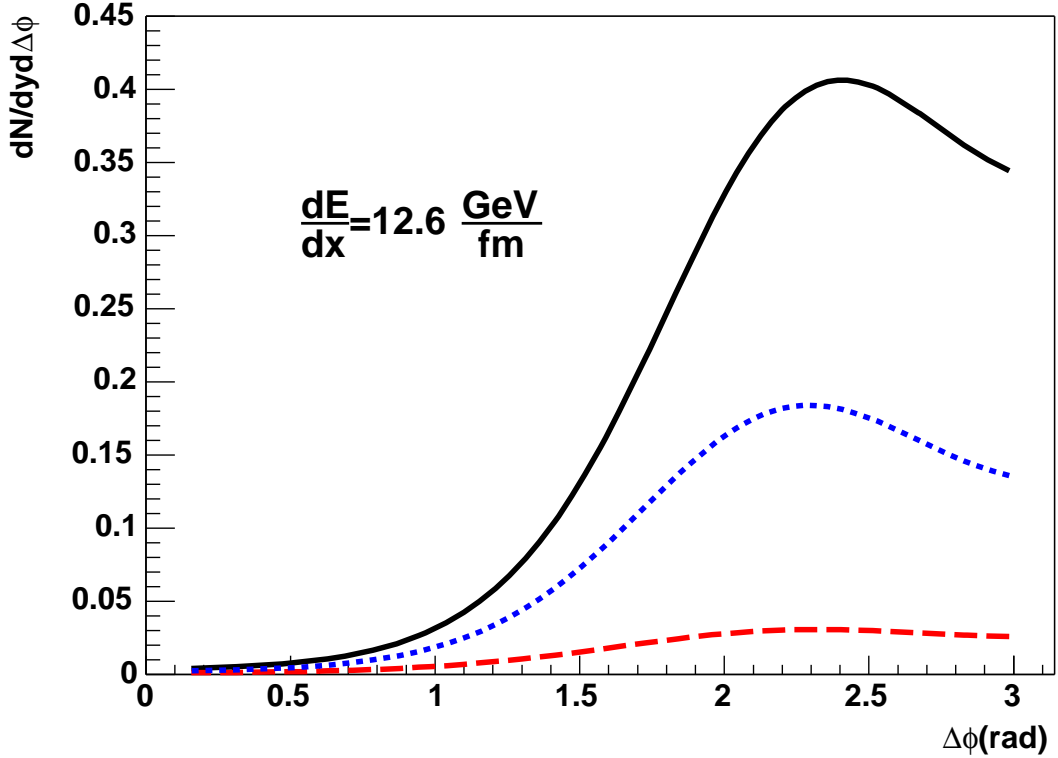}
\end{center}
\caption{
\label{Sdep}
Associate yield dependence on source size $\sigma$, for fixed energy loss
$dE/dx=63 T^2$, $\Gamma_s=0.1/T$, $t_j=8/T$, $t_f=10/T$ and $10 T<p_{t}<20 T$.
The three curves are for $\sigma=1/T$,
$\sigma=0.75/T$ and $\sigma=0.5/T$ for solid, dashed and dotted respectively.
}
\end{figure}

In figure \ref{Sdep} we study the source size dependence of the 
number of associated particles to the jet $dN/dyd\Delta \phi$ 
in an interval of transverse momentum $10 T_f<p_T<20 T_f$ for the mid rapidity
region y=0 for fixed values of the energy loss ($dE/dx=63 T^2$) 
and the jet energy ($t_j=8/T$) at a freeze out time of $t_f=10/T$.
 We observe that the smaller the source size the smaller the 
signal in the azimuthal dihadron correlations. This dependence is due mainly
to 
the viscosity. When the source size is comparable to the sound attenuation 
length, viscous effects reduce the amplitude very fast. For larger sources,
the reduction of the amplitude is delayed till times of the order of
\be
t_r\approx \sigma^2/\Gamma_s ~.
\ee
We also notice
that as the size grows, the peak in the correlation shifts slightly toward
$\pi$. This is because the interference of sound waves leads to the Mach angle 
 for propagation and observation times
 larger than the typical size of the object (as an extreme example consider
a large object that moves an infinitesimal distance at supersonic velocity;
it is clear then that it only can generate spherical waves).

\begin{figure}[t]
\begin{center}
\includegraphics[width=10cm,angle=-90]{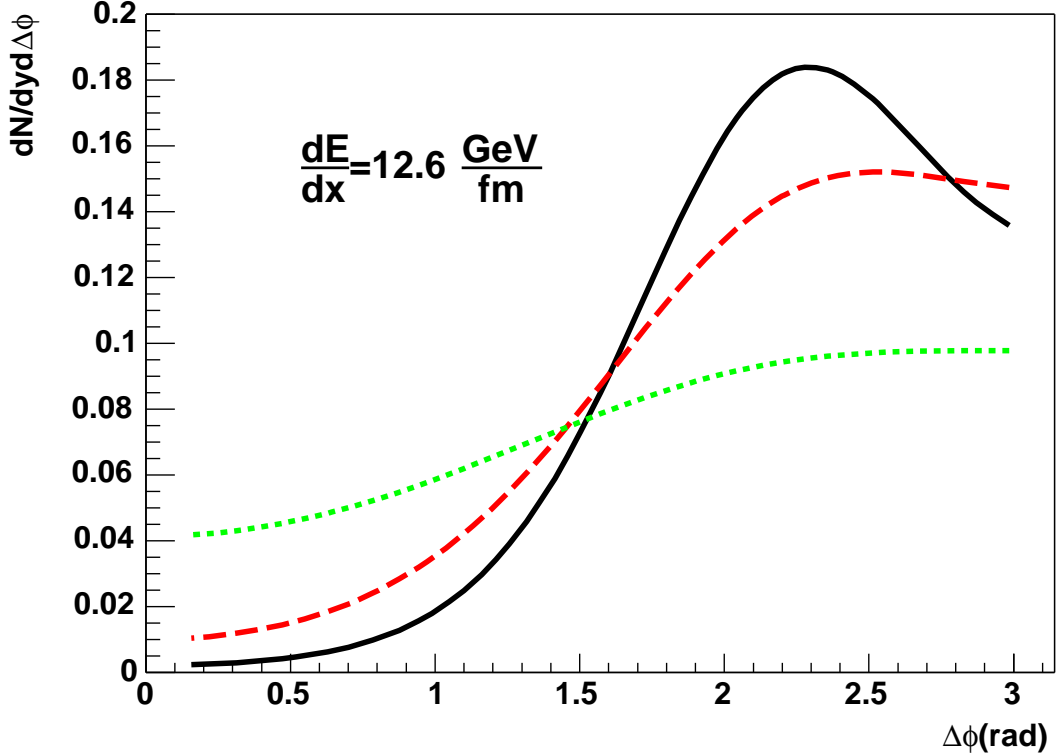}
\end{center}
\caption{
\label{Gsdep}
Associate yield dependence on viscosity for fixed source size $\sigma=0.75/T$
and energy loss
,  $dE/dx=63 T^2$, $t_j=8/T$, $t_f=10/T$ 
  and $10 T<p_{t}<20 T$. The three curves are for 
$\Gamma_s=0.1/T$, ($5 \times$) $\Gamma_s=0.2/T$,
 ($20 \times$) $\Gamma_s=0.5/T$
 for solid, dotted and dashed respectively }
\end{figure}

In figure \ref{Gsdep} we show the dependence on the sound attenuation length
$\Gamma_s$ for 
$\sigma=0.75/T$
and $dE/dx=63 T^2$ for three different values of the viscosity 
$T~\Gamma_s=0.1,0.2,0.5$ for solid, dotted and dashed lines respectively 
(note that the two last ones are rescaled up by a factor 10 and 20 
respectively). As in the case of the amplitude, the 
correlation function is very sensitive to the value of the viscosity. 
By changing the viscosity by a factor of 2 the whole correlation structure 
almost disappears and gets reduced by an order of magnitude. Increasing
the viscosity by a factor 5 leads to hardly no signal in the dihadron 
azimuthal distribution. This is because the dissipative effects induce
a reduction of the amplitude of the fields.

\begin{figure}[t]
\begin{center}
\includegraphics[width=10cm,angle=-90]{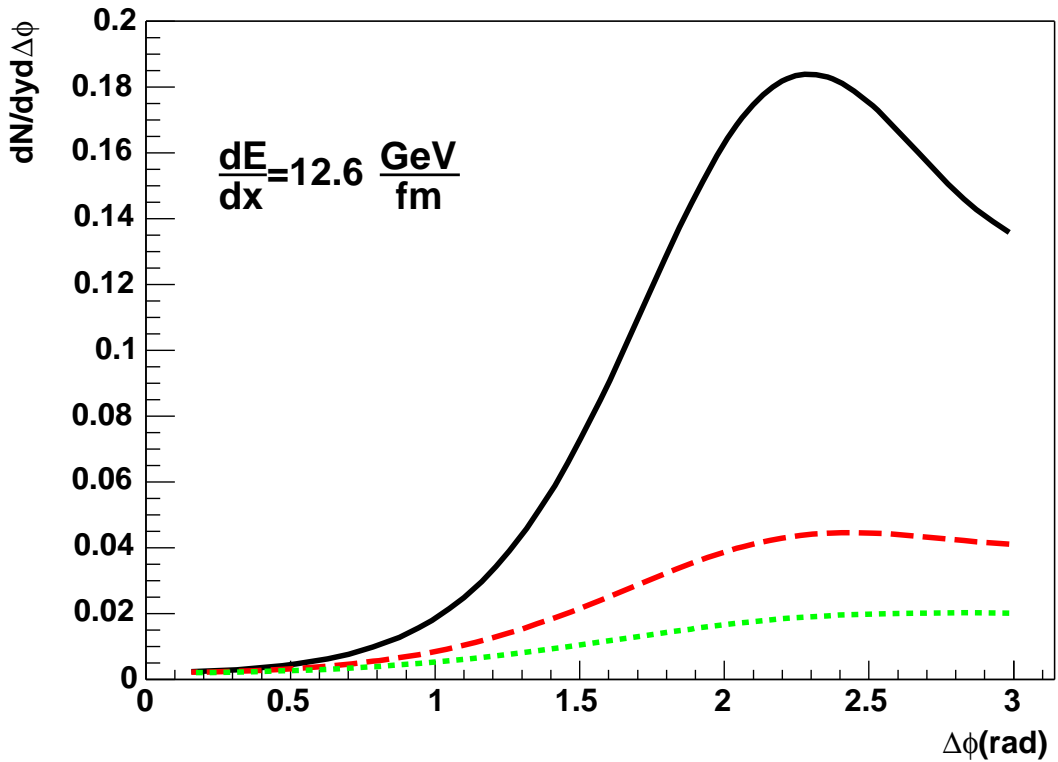}
\end{center}
\caption{
\label{tddep}
Associate yield dependence on the jet life time $t_j$ 
for fixed source size $\sigma=0.75/T$
energy loss
,  $dE/dx=63 T^2$, viscosity $\Gamma_s=0.1/T$, $t_j=8/T$, $t_f=10/T$ 
 and $10 T<p_{t}<20 T$.
 The three curves are for 
$t_j=8/T$, $t_j=7/T$, $t_j=6/T$,
 for solid, dotted and dashed respectively }
\end{figure}

In figure \ref{tddep} we study the dependence on the energy of the jet
though the jet life time for a fixed energy loss. We also fixed the source
size $\sigma$ and the viscosity $\Gamma_s$. We again observe a strong 
dependence of the correlation strength on the energy of the jet. This 
is in a sense due to the dependence on the viscosity. As the jet has
less energy it gets absorbed earlier in the medium and thus the 
sonic disturbances have to propagate longer before freeze out. Thus, the
attenuation of the waves leads to very strong suppression of the correlation
signal. The strong dependence on the path length complicates the 
interpretation of the large angle correlations due to the conical flow, 
as it seems to filter out a tight range of jet energies, reducing the 
total amplitude. However
this dependence on the path may not be directly extrapolate to
the real expanding case, as in such situation, the energy loss and the 
viscosity will depend on time. Thus, even though longer times are required
to observe a particle that is absorbed earlier, the large values of the
amplitude and the smaller value of the viscosity at early times leads to
an enhancement on the final observation. Thus, detailed studies in a 
real case should be performed before conclusions on the path length
dependence.

By the previous studies of the different parameters we want to show
that the appearance of conical flow seems to be very fragile, given
the large changes observed in the correlation for small changes of the
different characteristic scales
lead to completely different correlation strengths and shapes.
 However, if the interpretation of the large angle correlation
as coming from conical flow survives further tests, this strong dependence
reveal that this prove seems to be very
 sensitive to the 
details of the dissipative processes in the medium.
This strong dependence leads us to have some hope to use this prove in 
order to learn about the transport properties on the medium. The situation
is however unfortunate at the moment due to the freedom we have in the
parameters. If we could constrain the source size by some microscopic
model, the energy loss would constrain the amplitude of the wave, leaving
us with a very sensitive prove of the viscosity. Unfortunately such 
constraints are not available at the moment.

\section{Conclusions}
In this paper we have extended our studies in Ref.~\cite{CST} 
of the linearized hydrodynamic equations which describe
the interaction of jets with 
quark gluon plasma. We have clarified the origin
of the two hydrodynamic modes that we called sound and diffuson
in Ref.~\cite{CST} and related them to vortex free and rotational
solutions.

We then used the relativistic analogue potential flow
to find the governing equations of these 
modes in an expanding background.
For a static background, we subsequently used
these equations to systematically study the
interaction of a jet with the medium.
Generally the jet can excite both the sound and the diffusion
modes. 
The entropy produced by the jet-medium interaction 
fixes the strength of the diffusion mode relative 
to the sound mode. This entropy 
is localized in a narrow wake along the direction of the 
jet.

The equations for the sound wave are given by Eqs.~(\ref{def_pot}), and (\ref{phi_fluid}). The unknown function in this formula, $dF/dx$,  is
related to the momentum transfered to the sound wave -- see \eq{dedt}. 
Similarly the flow fields in the wake are given by 
Eq.~(\ref{Rgaus}), and the amplitude $A$ in this formula is 
fixed by the momentum transfered to the wake -- see \eq{Pwake}.
This momentum transfer is in turn related by \eq{Swake} 
to the total entropy
produced by the jet. In summary, by specifying 
the total rate of energy loss and entropy production the flow 
fields at large distances are determined.

Regardless of the mechanism of excitation, a supersonic 
jet in a static fluid leads to
 flow at the Mach angle. However, 
the different mechanisms of excitation may or may not lead
to visible peaks in the final particle spectra. 
If the  jet-medium interaction  produces significant entropy,
the wake contribution obscures the Mach angle in the final spectrum.
Only very strong jet-medium interaction which does not produce significant 
entropy can yield hydrodynamic fields that give
peaks in  the azimuthal correlation functions at the Mach angle,   $\Delta \phi \approx \pi - 1.2 ~\mbox{rad}$.
For example, for an isentropic jet-medium interaction with a large 
energy loss $dE/dx \approx 12 ~\mbox{GeV/fm}$ in a static medium, the resulting flow 
produces
azimuthal distributions of the same order of magnitude as those
observed by PHENIX and STAR \cite{phenix_peaks,star_peaks}.
Unfortunately, for  moderate energy loss,
 $dE/dx=2 ~\mbox{GeV/fm}$, the flow is unable to qualitatively reproduce
 the data. However,
we argued previously that the expansion 
may amplify the sound wave and ultimately reduce the required $dE/dx$ \cite{CS}.
We have also shown that for the large $dE/dx$,
the  $p_T$ dependence of the correlation function has some of the qualitative
features of the experimental measurements.

If the jet-medium interaction does not produce significant 
entropy,
the observation of peaks in the angular distribution is possible, but it
is very sensitive to the characteristic scales involved. 
Systematic study of the viscosity, the jet-medium interaction
time, the  freeze out time,  and the source size $\sigma$, 
reveals that the prediction of a Mach cone from linearized
hydrodynamics is quite fragile and depends on many of the microscopic details 
of the interaction of the jet with the medium. Further 
work is needed to clarify these microscopic details before
a strong conclusion can be reached.
\\
\\
\noindent {\bf Acknowledgments.} 
This work was partially supported by the 
 Department of Energy (U.S.A.) under grants DE-FG02-88ER40388
and DE-FG03-97ER4014.

\newpage

\appendix
\section {About the equations of the potential}
\subsection{\label{din_eq} Derivation of the dynamic equation}
In this section we derive a nonlinearequation for the potential
introduced in section 4.
As in the case of irrotational flow for ideal hydrodynamics,
the only non trivial equation is the 
 entropy 
continuity equation \eq{entopy_continuity}
the equation for the potential arises from re expressing \eq{entopy_continuity}
 in terms of $\phi$.
Let us start by writing \eq{entopy_continuity} as
\be
\label{D1}
\partial_{\mu} \frac{s}{T} \partial^{\mu} \phi + 
\frac{s}{T} \partial_{\mu}  \partial^{\mu} \phi=0~.
\ee 
As for a baryon free fluid the speed of sound can be expressed as 
\be
c^2_s=\frac{s}{T} \frac{dT}{ds} ~,
\ee
we can express the first derivative on \eq{D1} in terms of derivatives of
the temperature as
\be
\partial_{\mu} \frac{s}{T} = \frac{s}{T} 
                          \left(\frac{1}{c^2_s}-1 \right) \frac {1}{2 T^2}
                          \partial_{\mu} T^2 ~.
\ee
Now we note that from the definition of 
$\phi$ we obtain
\be
\label{Tofphi}
T^2=\partial_{\mu}\partial^{\mu} \phi ~.
\ee
Dropping the factor $s/T$ and multiplying by $T^2$, we obtain 
the following nonlinear equation equation for $\phi$
\be
\label{potential_nl}
\left(\frac{1}{c^2_s}-1\right) 
\partial_{\mu}\phi  \partial_{\nu}\phi  \partial^{\mu}\partial^{\nu}\phi+
\partial_{\mu}\phi  \partial^{\mu}\phi  \partial_{\nu}\partial^{\nu}\phi=0 ~.
\ee
For a baryon free fluid, the hydrodynamic equation reduces to the 
energy momentum conservation. Thus,  
in the case where the initial conditions for ideal hydrodynamics 
do not introduce any vorticity, the equation derived is totally equivalent
to solving the standard system of equation of conservation of the stress energy 
tensor.

\subsection {\label{Bj_pot} Bjorken solution as potential flow.}
As an example of potential flow let us look for a boost invariant solution 
of hydrodynamics with no transverse expansion \cite{Bj}. We will assume
that the initial conditions do not introduce any vorticity, and thus, 
all the properties of the flow field can be derived form a potential $\phi$.
As $\phi$ is a scalar, the requirement of boost invariance imposes that
$\phi$ is a function of the proper time $\tau=\sqrt{t^2-z^2}$ only.
For this potential, the nonlinear differential equation \eq{potential_nl}
takes the form
\be
\label{Bj_nl}
\left(\frac{d}{d\tau}\phi \right)^2
\left( \frac{1}{c_s^2}\frac{d^2}{d^2\tau}\phi-\frac{1}{\tau}\frac{d}{d\tau}\phi\right)
=0~.
\ee
Using the relation between the temperature and the potential \eq{Tofphi} 
we obtain T as a function of $\tau$
\be
T=\frac{d}{d\tau}\phi(\tau) ~.
\ee
Thus, \eq{Bj_nl} becomes a first order equation for the temperature
\be
\frac{d}{d\tau} T +\frac{c^2_s}{\tau} T=0~,
\ee
which, by means of the equation of state leads to
\be
\frac{1}{s}\frac{d}{d\tau}s + \frac{1}{\tau}=0 ~,
\ee
the solution of which is $s \propto 1/\tau$, the well know Bjorken
solutions \cite{Bj}. 
The velocity field can also be readily calculated from the potential
 From its  definition,   
$T u_{\mu}=\partial_{\mu} \phi$,  we obtain that
the the velocity field is  $u^{\tau}=1$ or 
\be
u^{\mu}=\frac{1}{\tau}(t,0,0,z) ~,
\ee
that coincides again  with the result due to Bjorken \cite{Bj}. 

\section{\label{matching} Far field from summing infinitesimal disturbances}
We now want to connect our previous solution on \cite{CST} with our general
expression \eq{phi_fluid} for the perturbed fields in the ideal case.
The goal is to show that both ways of calculating give the same expression
in the region far from the fluid. However, extra terms from those in 
\eq{phi_fluid} appear that are only significant in the region close
to the jet. 

As argued along the text, the potential flow describes sound waves, that are
also described for the system of equations \eq{eglsys}. 
Setting $\eta=0$, the system of equations ca be written as the wave equation
for $\epsilon$. We will consider the solution for an infinitesimal disturbance
that happens in an time interval $dt_0$ around $t_0$ that leads
spherically symmetric disturbance
\be
\label{sph_ini}
\epsilon_{dt_0}(t_0,{\bf x })=e_{0}({\bf R}_{t_0}) ~ ,\\
g_{dto}(t_0,{\bf x })=g_{0}({\bf R}_{t_0})~,
\ee
where ${\bf R}_{t_0}=\sqrt{(x+t_0)^2+\rho^2}$ for a jet moving at the speed of light
along the $-x$ direction.The total disturbance is obtained by
the addition of all the infinitesimal disturbances.

The general solution for the spherically symmetric wave takes the form
\be
\label{inf_sol}
\epsilon_{t_0}=\frac{1}{R_{t_0}}\left( f_{-}\left(R_{t_0}-c_s(t-t_0)\right) 
+  f_{+}\left(R_{t_0}+c_s(t-t_0)\right) \right)
\ee
The values of the two functions are set by the initial condition
\eq{sph_ini} at $t_0$. 
 As the incoming wave $f_+$ is only
important in the region of the order of the source size
we can neglect it in the far field. 

From the definition of the potential in the static case, 
the perturbed temperature can be as $T'=\del_t\varphi$
and thus, by means of the equation of state, we 
obtain the potential
\be
\varphi_{dt0}=\frac{1}{R_{t_0}}\left( F\left(R_{t_0}-c_s(t-t_0)\right)\right) ~,
\ee
where $ \dot F=c^2_s f_{-}/s$ where $ \dot F $ denotes time 
derivative. The addition of all the  infinitesimal disturbances
leads to 
\be
\label{sph_sum}
\varphi=\int^{\infty}_{-\infty} dt_0 \theta(t-t_0) 
        \frac{F\left(R_{t0}-c_s(t-t_0)\right)}{R_{t0}} ~.
\ee

 The solution \eq{sph_sum} can be rewritten as \eq{phi_fluid} by
introducing a $\delta-$function as
\be
F\left(R_{t0}-c^2_s(t-t_0)\right)=
                     \int^{\infty}_{-\infty} d\xi \frac{c_s}{v} 
                                           F\left(-\frac{c_s}{v} \xi \right)
                    \delta \left(R_{t0}-c_s(t-\xi/v-t_0)\right)
\ee
The integration of over $t_0$ can be done now by solving the delta function.
This is standard and after this solution, we can express
\be
\frac{\delta \left(R_{t0}-c_s(t-\xi/v-t_0)\right)}{R_{dt0}}=
\frac{\delta\left(t_0-t_i\right)}{c_s\left((x+vt-\xi)^2-\beta^2\rho^2\right)^{1/2}}
\ee
 with $\beta$ defined
 in \eq{beta_def}, and $t_i$ is the solution of the $\delta-$function.
\be
\label{ti_sol}
t_i-t+\frac{\xi}{v}=\frac{v}{v^2-c^2_s}
                    \left(-\left(x+vt-\xi\right)\pm
                     \sqrt{\left(x+vt-\xi\right)^2-\beta^2\rho^2}\right)~.
\ee
The existence of solutions for the $\delta-$function imposes a constraint
in the $\xi$ integration such that the radicand in \eq{ti_sol} is greater
than zero.
The $\theta-$function in \eq{sph_sum} constraints further this integration.
Imposing $t>t_i$ leads
to two different integration regions:

\begin{enumerate}
\item  $x+vt > c^2_s/v^2 (x+vt-\beta \rho)$, that includes the region inside
of the Mach cone. In this case requiring that $t>t_0$ leads to the field
\be
\label{incone}
\varphi
=\frac{1}{v}\left( 2 \int^{x+vt-\beta \rho}_{-\infty} d\xi 
                   - \int^{-v R_t /c }_{-\infty} d\xi \right) 
       \frac{ F\left(-\frac{c_s}{v} \xi \right)}{\left((x+vt-\xi)^2-\beta^2\rho^2\right)^{1/2}}
\ee
\item  $x+vt <c^2_s/v^2 (x+vt-\beta \rho)$, that corresponds to the field out of the cone
The field is then
\be
\varphi=\frac{1}{v}
\int^{-v R_t /c }_{-\infty} d\xi
\frac{ F\left(-\frac{c_s}{v} \xi \right)}{\left((x+vt-\xi)^2-\beta^2\rho^2\right)^{1/2}}
\ee

\end{enumerate}


\begin{thebibliography}{99}

\bibitem{jetquenching} 
K. ~Adcox {\it et al.} (PHENIX), {\it Phys. Rev. Lett.} {\bf 88} (2002)
022301
\\
C. ~Adler {\it et al.} (STAR),   {\it Phys. Rev. Lett.} {\bf 89} (2002)  202301
\\
S.~ S.~ Adler {\it et al.} (PHENIX), {\it Phys. Rev. Lett.} {\bf 91} (2003) 072301
\\
J. ~Adams  {\it et al.} (STAR),  {\it Phys. Rev. Lett.} {\bf 91}  (2003) 172302

\bibitem{early}
J.~ D.~ Bjorken,
FERMILAB-PUB-82-059-THY

\bibitem{Gyulassy_losses}
X. ~N. ~Wang, M. ~Gyulassy  and M. ~Plumer, 
{\it Phys. Rev.} D {\bf 51} (1995) 3436
\\
M.~Gyulassy, I.~Vitev, X.~N.~Wang and B.~W.~Zhang,
arXiv:nucl-th/0302077.

\bibitem{Kovner:2003zj}
  A.~Kovner and U.~A.~Wiedemann,
  arXiv:hep-ph/0304151.

\bibitem{Dok_etal}
R. ~Baier, Y.~ L.~Dokshitzer,A.~ H.~ Mueller,S. ~Peigne and D.~Schiff, 
{\it Nucl. Phys.} B  {\bf 483} (1997) 291 
\\
R. ~Baier , Y. L. ~Dokshitzer, A. ~H. ~Mueller  and D. ~Schiff,
{\it JHEP} {\bf 0109} (2001) 033


\bibitem{SZ_dedx}
E. ~V. ~Shuryak and I.~ Zahed  
arXiv:hep-ph/0406100


\bibitem{xnwang_workshop} 
Wang X N, 2004 [arXiv:nucl-th/0405017]

\bibitem{CST}J.~Casalderrey-Solana, E.~V.~Shuryak and D.~Teaney,
{\it J. Phys. Conf. Ser.} {\bf 27} (2005) 22, [arXiv:hep-ph/0411315]


\bibitem{Stocker} H. Stoecker, [arXiv:nucl-th/0406018]


\bibitem{sQGP}
  E.~Shuryak,
  Prog.\ Part.\ Nucl.\ Phys.\  {\bf 53} (2004) 273 
  [arXiv:hep-ph/0312227].

\bibitem{Baymetal}
G.~Baym, H.~Monien, C.~J.~Pethick and D.~G.~Ravenhall,
Phys.\ Rev.\ Lett.\  {\bf 64} (1990) 1867.


\bibitem{AMY6}
P.~Arnold, G.~D.~Moore and L.~G.~Yaffe,
JHEP {\bf 0305}, 051 (2003)
[arXiv:hep-ph/0302165].

\bibitem{Derek_visc} D. Teaney,  {\it Phys.Rev.} C {\bf 68} (2003)
   034913

\bibitem{phenix_peaks} 
S.~S. ~Adler {\it et al.} (PHENIX Collaboration), [arXiv:nucl-ex/0507004]   
\bibitem{star_peaks} 
J~. Adams {\it et al.} (STAR Collaboration), {\it Phys. Rev. Lett.} {\bf 95}  
(2005) 152301 

\bibitem{RM} J. Ruppert and B. Muller, {\it Phys. Lett.} {\bf B} 618 123,2005

\bibitem{Dremin} I. M. Dremin,  [arXiv:hep-ph/0507167] 

\bibitem{MWK} V. Koch, A. Majumder, X. N. Wang, [arXiv:nucl-th/0507063] 

\bibitem{Vitev} I. Vitev, [arXiv:hep-ph/0501255] 


\bibitem{S_w} L. M. Satarov, H. Stoecker, I. N. Mishustin, {\it Phys. Lett.} {\bf B } 627 64,2005

\bibitem{Renk} T. Renk, J. Ruppert,  [arXiv:hep-ph/0509036] 


\bibitem{Heinz}
A. ~K. ~Chaudhuri and  ~U. ~Heinz, [arXiv:nucl-th/0503028] 

\bibitem{MAMPT}
G. ~L. ~Ma {\it et al.}, [arXiv:nucl-th/0601012]

\bibitem{AMPT}
Z. ~W. ~Lin {\it et al.}, {\it Phys. Rev.} C {\bf 72} (2005)  064901 


\bibitem{CS}
J.~ Casalderrey-Solana, E.~V.~ Shuryak, [arXiv:hep-ph/0511263]

\bibitem{Baier}
R. ~Baier, {\it Nucl. Phys.} A {\bf 715} (2003)   209 
 
\bibitem{karsch}
F.~ Karsch, {\it Lect. Notes Phys.} {\bf 583} (2002)   209

\bibitem{visc_bound} G. ~Policastro, D. ~T.~ Son and A.~ 0.~ Starinete,
{\it Phys. Rev. Lett.} {\bf 87} (2001) 081601


\bibitem{landau}
L.~ D.~ Landau and E.~ M.~ Lifshitz, {\it Fluid Mechanics}

\bibitem{Taub}
A.~H.~Taub, Ann. Rev. Fluid. Mech. {\bf 10} (1978)  301

\bibitem{Bj}
J.~D.~Bjorken, {\it Phys. Rev} D {\bf 27} (1983) 140 

\bibitem{Bilic}
N. ~Bilic, [arXiv:gr-qc/9908002]

\bibitem{viser} 
C. ~Barcelo, S. ~Liberati,  M.~ Visser,  [arXiv:gr-qc/0505065]

\bibitem{VPF}
D.~D.~Joseph, J. Fluid. Mech. {\bf479} (2003) 191

\bibitem{cf} 
F.~ Cooper and G.~ Frye, {\it Phys. Rev.} D {\bf 27} 140.

\bibitem{EHSW}
K.~ J.~ Eskola, H.~ Honkanen, C.~A.~Salgado, U.~ A.~ Wiedemann, 
{\it Nucl. Phys.} A {\bf 747} (2005) 511 


\end{thebibliography}
\end{document}